\documentclass[12pt]{article} %
\usepackage{latexsym}
\usepackage{amsfonts}
\usepackage{amsmath}
\usepackage{graphicx}
\usepackage{amssymb}

\numberwithin{equation}{section}

\def\tc{{\tt c}}
\def\td{{\tt r}}

\def\tR{{\tt R}}

\def\tS{{\tt S}}
\def\ts{{\tt s}}

\def\ttt {{\tt t}}

\def\tU{{\tt U}}

\def\t0{{\tt 0}}

\def\vphi{\varphi}

\def\orho{{\overline\rho}}

\def\om{\omega}
\def\oom{\overline\omega}

\def\bom{{\mbox{\boldmath$\omega$}}}
\def\obom{\overline\bom}

\def\0bom{{\bom}^0}
\def\0obom{{\obom}^0}

\def\0nbom{{\bom}_{n,0}}
\def\n*bom{{\bom}^*_{(n)}}
\def\wt{\widetilde}

\def\oom{\overline\om}

\def\Om{\Omega}
\def\oOm{\overline\Om}
\def\uOm{\underline\Om}
\def\bOm{\mbox{\boldmath${\Om}$}}
\def\obOm{\overline\bOm}

\def\Gam{\Gamma}
\def\gam{\gamma}
\def\Lam{\Lambda}
\def\Lamc{\Lam^{\rm c}}

\def\ozeta{{\overline\zeta}}

\def\bbC{\mathbb C}

\def\fB{\mathfrak B}

\def\fF{\mathfrak F}

\def\fK{\mathfrak K}

\def\fW{\mathfrak W}

\def\rd{\rm d}
\def\rmm{\rm m}

\def\dist{\rm{dist}}

\def\cC{\mathcal C}

\def\cH{\mathcal H}

\def\cW{\mathcal W}

\def\ocW{\overline\cW}
\def\ucW{\underline\cW}

\def\bbC{\mathbb C}
\def\bbP{\mathbb P}
\def\obbP{\overline\bbP}
\def\ubbP{\underline\bbP}

\def\bbR{\mathbb R}

\def\bx{\mathbf x}

\def\ux{\underline x}

\def\by{\mathbf y}
\def\uy{\underline y}

\def\ox{\overline x}

\def\bz{\mathbf z}

\def\dist{\textrm{dist}}
\def\diy{\displaystyle}
\def\ov{\overline}

\def\ux{\underline x}
\def\uy{\underline y}

\def\om{\omega}
\def\oom{\ov\om}

\def\wh{\widehat}

\def\ucW{\underline\cW}

\def\rd{{\rm d}}

\def\tr{{\rm{tr}}}
\def\rtr{{\rm{tr}}}
\def\orZ{{\overline{\rm Z}}}

\begin{document}
\title{\bf FK-DLR states\\
of a quantum bose-gas\\
with a card-core interaction}
\author{\bf Y. Suhov~$^{1}$, M. Kelbert$^{2}$}
\vspace{1mm}\maketitle {\footnotesize

\noindent $^1$ Statistical Laboratory, DPMMS, University of Cambridge, UK;\\
Department of Statistics/IME,  University of S\~ao Paulo, Brazil;\\
IITP, RAS, Moscow, Russia\\
E-mail: yms@statslab.cam.ac.uk

\noindent $^2$ Department of Mathematics, Swansea University, UK;\\
Department of Statistics/IME, University of S\~ao Paulo, Brazil\\
E-mail: M.Kelbert@swansea.ac.uk}

\begin{abstract}

This paper  and its sequels, \cite{SKS} and \cite{SS}, continue the works \cite{KS1}, \cite{KS2} 
and \cite{KSY2}. We pursue two lines of study: (i) a definition of an 
infinite-volume quantum Gibbs state for various types of quantum bosonic 
systems, and (ii) its justification, which we have chosen to be the shift-invariance 
property for two-dimensional Bose-gas; cf. \cite{SKS}. In \cite{SS} the above results are
established for non-negative interaction potentials.  

We consider a  particle system (a quantum gas) in $\bbR^d$. 
The kinetic energy part 
of the Hamiltonian is the standard Laplacian (with a Dirichlet's boundary condition
at the border of a `box').   
The particles interact with each other through a two-body 
finite-range potential $V$ depending on the distance between them
and featuring a hard core of diameter $\td>0$. We introduce a
class of so-called FK-DLR functionals containing all limiting Gibbs 
states of the system. In the next paper we will prove that any FK-DLR functional 
is shift-invariant, regardless of whether it is unique or not.
\vskip.1truecm

\end{abstract}
\vskip 1 truecm

\section{Introduction. Limit-point Gibbs states and reduced density matrices}

\medskip
 
{\bf 1.1. The local Hamiltonian.} The object of study in this paper is a 
quantum Bose-gas in a Euclidean space $\bbR^d$.
The starting point of our analysis is a self-adjoint $n$-particle Hamiltonian,
$H_{n,\Lam}$, of the system in a finite `box' $\Lam$. Typically, $\Lam$ is
represented by a cube $[-L,+L]^{\times d}$, of size $2L>0$, centered at the 
origin and with edges parallel to the co-ordinate axes. (Other types of bounded 
domains in $\bbR^d$ can also be incorporated.) The operator $H_{n,\Lam}$ is 
given by
$$\begin{array}{l}
\diy\left(H_{n,\Lam}\phi_n\right)\left(\ux_1^n\right)
=-\frac{1}{2}\sum\limits_{1\leq j\leq n} 
\left(\Delta_j\phi_n\right)\left(\ux_1^n\right)\\
\qquad +\sum\limits_{1\leq j<j'\leq n}V\left(\left|x(j)-x(j')\right|\right)
\phi_n\left(\ux_1^n\right),\;\;\ux_1^n=\{x(1),\ldots ,x(n)\}\in\left(\bbR^d
\right)^n\end{array}\eqno (1.1.1)$$
and acts on functions $\phi_n\in {\rm L}_2^{{\rm{sym}},\td}(\Lam^n)$. 
Here ${\rm L}_2^{{\rm{sym}},\td}
(\Lam^n)$ stands for the subspace in the Hilbert space ${\rm L}_2(\Lam^n)=
{\rm L}_2(\Lam )^{\otimes n}$ formed by symmetric functions of variables $x(j)$,
$1\leq j\leq n$, constituting the argument $\ux_1^n$, which vanish whenever
$$\min\;\left[\left|x(j)-x(j')\right|_{\rm{Eu}}:\;1\leq j<j'\leq n\right]<\td.$$
(Here and  below, $|x|_{\rm{Eu}}$, or briefly $|x|$, stands for the Euclidean 
norm of $x\in\bbR^d$ whereas $|x|_{\rmm}$ denotes the max-norm.)   
Parameter $\td>0$ is fixed and represents the diameter of the hard core
(see below). It is convenient to denote 
$$\begin{array}{l}\Lam^n_\td=\Big\{\ux_1^n=(x(1),\ldots ,x(n))\in\Lam^n:\\
\qquad\qquad
\min\;\left[\left|x(j)-x(j')\right|:\;1\leq j<j'\leq n\right]\geq\td\Big\}\end{array}\eqno (1.1.2)$$
and identify  ${\rm L}_2^{{\rm{sym}},\td}(\Lam^n)$ with   
${\rm L}_2^{{\rm{sym}}}(\Lam^n_\td)$, the Hilbert space of square-integrable
symmetric functions $\phi_n\left(\ux_1^n\right)$ with support in $\Lam^n_\td$.

Operator $\Delta_j$ in (1.1.1) acts as a Laplacian in the variable $x(j)$.
Further, $V:\,r\in [\td,+\infty )\mapsto V(r)\in\bbR$ is a $C^2$-function 
with a compact support, describing a two-body interaction potential depending 
upon the distance between particles. The value
$$\tR=\inf\,\left[r>0:\;V({\wt r})\equiv 0\;\hbox{ for }\;{\wt r}\geq r\right]\eqno (1.1.3)$$
is called the interaction radius (or the interaction range). Formally speaking,
we set: $V(r)=+\infty$ for $0\leq r<\td$, conforming with the hard-core assumption.
We also set:
$$-{\ov V}=\min\,\big[V(r):
\;\td\leq r\leq \tR\big],\eqno (1.1.4)$$
with ${\ov V}=0$ for $V\geq 0$, and
$${\ov V}^{\,(1)}=\max\,\big[\left| V'(r)\right|: \td\leq r\leq\tR\big]
.\eqno (1.1.5)$$

\medskip

The above assumptions upon the two-body potential $V$ are valid throughout the 
whole paper.

\medskip 

For $n=1$, the sum $\sum\limits_{1\leq j<j'\leq n}$ in Eqn (1.1.1) is 
suppressed, and 
$H_{n,\Lam}$ is reduced to $\diy -\frac{1}{2}\Delta$ in $\Lam$. For $n=0$,
we formally define $H_{0,\Lam}=0$. In general, the term
$\diy-\frac{1}{2}\sum\limits_{1\leq j\leq n}
\left(\Delta_j\phi_n\right)\left(\ux_1^n\right)$ represents the kinetic 
energy part in the Hamiltonian, and the term 
$\sum\limits_{1\leq j<j'\leq n}V\left(\left|x(j)-x(j')\right|\right)$ 
the potential energy (as an operator, it is given as multiplication
by this function). Note that if $n$ is large enough (viz., $n\pi\Gam ((d+1)/2)\td^d \geq (2L)^d
=\hbox{ volume of\;}
\Lam$) then the expression for $H_{n,\Lam}$ formally becomes infinite;
consequently, we only care about the values of $n$ such that the set 
$\Lam^n_\td\neq\emptyset$. 

Operator $H_{n,\Lam}$ is determined by a boundary condition. More precisely, 
it is initially defined by the RHS of Eqn (1.1.1) as a symmetric operator 
on the set of C$^2$-functions $\phi =\phi_n$ with the support  in the interior of $\Lam^n_\td$.
A self-adjoint extension of this symmetric operator is determined by boundary conditions.  
On the set $\partial^{\rm{hc}}\Lam^n_\td$: 
$$\begin{array}{l}\partial^{\rm{hc}}\Lam^n_\td=
\Big\{\ux_1^n=(x(1),\ldots ,x(n))\in\Lam^n_\td:\\
\qquad\qquad\min\;\left[\left|x(j)-x(j')\right|:\;1\leq j<j'\leq n\right]=\td\Big\}
\end{array}\eqno (1.1.6)$$ 
we take the Dirichlet boundary condition:
$$\phi (\ux_1^n)=0\;\hbox{ for }\;\ux_1^n\in\partial^{\rm{hc}}\Lam^n_\td.\eqno (1.1.7)$$

On the other hand, the cube $\Lam =\left[-L,L\right]^{\times d}$ 
also has an  `outer'  border $\partial\Lam=\{y\in\bbR^d:\;|y|_{\rmm}=L\}$.
Next, given $n\geq 2$, we define: 
$$\begin{array}{l}\partial^{\rm{out}}\Lam^n_\td
=\Big\{\ux_1^n=(x(1),\ldots ,x(n))\in\Lam^n_\td:\\
\qquad\qquad\qquad\max\left[\big|x(j)\big|_{\rmm}:\;1\leq j\leq n\right]=L\Big\}
.\end{array}\eqno (1.1.8)$$

Throughout the paper we consider Dirichlet's boundary condition on 
$\partial^{\rm{out}}\Lam^n_\td$:
$$\phi_n(\ux_1^n)=0,\;\;\ux_1^n\in\partial^{\rm{out}}\Lam^n_\td.\eqno (1.1.9)$$
Nevertheless, the methods of this work allow us to consider a broad class of boundary 
conditions, including Neumann's (zero of the normal derivative) and elastic (where 
a linear combination of the values of the function and its normal derivative vanishes); 
periodic boundary conditions
can also be included.  Considering various boundary conditions endeavors towards 
enhancing possible phase transitions; 
we intend to return to this question in a forthcoming work.

Under the above assumptions, operator $H_{n,\Lam}$ is self-adjoint, bounded 
from below and has a pure point spectrum. Moreover, $\forall$ $\beta\in (0,+\infty )$,
the Gibbs operator $G_{\beta ,n,\Lam}=\exp\,\left[-\beta H_{n,\Lam}\right]$ is a 
positive-definite trace-class operator in ${\rm L}_2^{{\rm{sym}}}(\Lam^n_\td)$. The trace
$$\Xi_{\beta,n}(\Lam ):={\tr}_{{\rm L}_2^{{\rm{sym}}}(\Lam^n_\td)}G_{\beta ,n,\Lam}
\in (0,+\infty )\eqno (1.1.10)$$  
is called the $n$-particle partition function in $\Lam$ at the inverse 
temperature $\beta$. When $n$ is large and $\Lam^n_\td$ becomes empty, we
set $G_{\beta ,n,\Lam}$ to be a zero operator
with $\Xi (\beta,n,\Lam )=0$. This allows us to work with the grand 
canonical Gibbs ensemble. Namely, $\forall$ $z\in (0,+\infty )$, 
the direct sum 
$$G_{\beta ,\Lam}=\operatornamewithlimits{\oplus}\limits_{n\geq 0}
z^nG_{\beta ,n,\Lam}\eqno (1.1.11)$$
determines a positive-definite trace-class operator in the bosonic Fock space 
$$\cH (\Lam )=\operatornamewithlimits{\oplus}\limits_{n\geq 0}
{\rm L}_2^{\rm{sym}}(\Lam^n_\td).\eqno (1.1.12)$$

The quantity
$$\Xi_{z,\beta }(\Lam ):=\sum_{n\geq 0}z^n\Xi (\beta,n,\Lam )
={\tr}_{\cH (\Lam )}G_{\beta ,\Lam}\in (0,+\infty ) \eqno (1.1.13)$$
is called the grand canonical partition function in $\Lam$ at fugacity
$z$ and the inverse temperature $\beta$. Further, the operator 
$$R_{\beta ,\Lam}=\frac{1}{\Xi_{z,\beta}(\Lam )}
G_{z,\beta,\Lam}\eqno (1.1.14)$$
is called the (grand-canonical) density matrix (DM) in $\Lam$;
this is a positive-definite operator in $\cH (\Lam )$ of trace $1$.
Operator $R_{z,\beta ,\Lam}$ determines the Gibbs state (GS),
i.e., a linear positive normalized functional $\varphi_{z,\beta ,\Lam}$ 
on the C$^*$-algebra $\fB(\Lam )$ of bounded operators in $\cH (\Lam )$:
$$\varphi_{z,\beta ,\Lam}(A)={\tr}_{\cH (\Lam )}\big(AR_{z,\beta ,\Lam}\big),
\;\;A\in\fB(\Lam ).\eqno (1.1.15)$$  

The next object of interest is the reduced DM (in short, the RDM), in cube
$\Lam_0\subset\Lam$ centered at a point $c_0=(\tc_0^1,\ldots ,\tc_0^d)$: 
$$\Lam_0=[-L_0+\tc_0^{1},\tc_0^{1}+L_0]\times\cdots\times [-L_0+\tc_0^{d},\tc_0^{d}+L_0].
\eqno (1.1.16)$$ 
We use the term RDM for the partial trace
$$R^{\Lam_0}_{z,\beta ,\Lam}=\tr_{\cH (\Lam\setminus\Lam_0)}R_{z,\beta,\Lam};
\eqno (1.1.17)$$
it is based on the tensor-product representation $\cH (\Lam )=\cH(\Lam_0)\otimes
\cH  (\Lam\setminus\Lam_0)$. Operator $R^{\Lam_0}_{z,\beta ,\Lam}$ acts in $\cH(\Lam_0)$, 
is positive-definite 
and has trace $1$. Moreover, the partial trace operation leads to an important
compatibility property for RDMs: if cubes $\Lam_1\subset\Lam_0\subset\Lam$ then
$$R^{\Lam_1}_{z,\beta ,\Lam} =\tr_{\cH (\Lam_0\setminus\Lam_1)}R^{\Lam_0}_{z,\beta ,\Lam} .
\eqno (1.1.18)$$
The mnemonic here is that the upper indices $\Lam_0$ and $\Lam_1$ indicate 
`volumes' that have been kept `free' of the partial trace. 

The main results of the present paper are valid $\forall$ $z,\beta\in (0,\infty )$
under the condition 
$$\orho:=z\exp\,(\beta {\ov V}\tR^d/\td^d)<1;\eqno (1.1.19)$$
cf. (1.1.4). Inequality (1.1.19) 
becomes $z\in (0,1)$ when the two-body potential $V\geq 0$.\footnote{Condition (1.1.19)
is needed to guarantee a `thermodynamic stability' of the system; apparently, it 
should not prevent phase transitions when the dimension $d\geq 2$.} Recall, $\td\in (0,\infty )$ is 
the diameter of the hard core and $\tR\in (\td ,\infty)$ the radius of interaction; cf. 
(1.1.2) and (1.1.4). To simplify the notation, we will omit the indices/arguments 
$z$ and $\beta$ whenever it does not lead to a confusion.
A straightforward generalization of the above concepts can be done by including
an external potential field induced by an external classical configuration (CC) $\bx (\Lamc)$ represented
by a finite or countable subset in the complement $\Lamc$ such that 
$$|\ox -\ox'|\geq\td\;\;\forall\;\;\ox,\ox'\in\bx (\Lamc)\;\hbox{ with }\;\ox\neq\ox' .\eqno (1.1.20)$$ 
(In fact, it is the 
intersection $\Lam^{(\tR)}\cap\bx (\Lamc)$ that will matter, where $\Lam^{(\tR)}=\{x\in\bbR^d:
\;\dist (x,\Lam)\leq\tR\}$. Here and below, $\dist$ stands for the Euclidean distance.)
Viz., the Hamiltonian $H_{n,\Lam |\bx (\Lamc )}$ is given by
$$\left(H_{n,\Lam |\bx (\Lamc )}\phi_n\right)\left(\ux_1^n\right)=\left(H_{n,\Lam}\phi_n\right)(\ux_1^n)
+\sum\limits_{1\leq j\leq n}\sum\limits_{\ox\in\bx (\Lamc )}V\left(\left|x(j)-\ox\right|\right)
\phi_n\left( \ux_1^n\right)\eqno (1.1.21)$$
and has all properties that have been listed above for  $H_{n,\Lam}$. This enables
us to introduce the Gibbs operators 
$G_{n,\Lam |\bx (\Lamc )}$ and $G_{\Lam |\bx (\Lamc )}$,
the partition functions $\Xi_{n}(\Lam |\bx (\Lamc ))$ and $\Xi (\Lam |\bx (\Lamc ))$,
the DM $R_{\Lam |\bx (\Lamc )}$, the GS $\varphi_{\Lam |\bx (\Lamc )}$ and
the RDMs $R^{\Lam_0}_{\Lam |\bx (\Lamc )}$ where $\Lam_0\subset\Lam$.\footnote{Although 
the Hamiltonian $H_{n,\Lam}$
and its derivatives $G_{n,\Lam }$, $G_{\Lam }$ and so on, are particular examples of 
$H_{n,\Lam |\bx (\Lamc )}$, etc. (with $\bx (\Lamc)$ being an empty configuration), we will 
now and again address this specific example individually, for its methodological significance.}
Viz.,
$$\begin{array}{c}G_{n,\Lam |\bx (\Lamc )}=\exp\,\left[-\beta H_{n,\Lam |\bx (\Lamc )}\right],\\ \;\\
 \Xi _n(\Lam |\bx (\Lamc )):={\tr}_{{\rm L}_2^{{\rm{sym}}}(\Lam^n_\td)}G_{n,\Lam |\bx (\Lamc )}
\in (0,+\infty ),\end{array}\eqno (1.1.22)$$
and
$$\begin{array}{c} G_{\Lam  |\bx (\Lamc )}=\operatornamewithlimits{\oplus}\limits_{n\geq 0}
z^nG_{n,\Lam  |\bx (\Lamc )},\\
\Xi (\Lam |\bx (\Lamc )):=\sum\limits_{n\geq 0}z^n\Xi_n(\Lam |\bx (\Lamc ))
={\tr}_{\cH (\Lam )}G_{\Lam |\bx (\Lamc )}\in (0,+\infty ) .
\end{array}\eqno (1.1.23)$$
In the case of an empty exterior $\bx (\Lamc )=\emptyset$, the 
argument $\bx (\Lamc )$ is omitted.

We conclude Section 1.1 with the following remark. The Fock spaces $\cH(\Lam )$ and 
$\cH (\Lam_0)$ (see (1.1.12)) can be conveniently 
represented as ${\rm L}_2(\cC_\td(\Lam ))$ and ${\rm L}_2(\cC_\td(\Lam ^0))$, respectively.
Here and below, $\cC (\Lam )$ denotes the collection of finite (unordered) subsets 
$\bx\subset\Lam$ (including the empty set) with the Lebesgue--Poisson measure
$$\begin{array}{l}\diy{\rd}\bx =\frac{1}{(\sharp\;\bx)!}\prod_{x\in\bx}{\rd}x,\;\;\sharp\;\bx <\infty\;\;
\hbox{(with $\diy\int\limits_{\cC (\Lam)}{\rd}\bx
=\exp\,[\ell (\Lam)]$}\\
\qquad\qquad\qquad\qquad\hbox{where $\ell$ is the Lebesgue measure on $\bbR^d$),}
\end{array}\eqno (1.1.24)$$
and $\cC_\td(\Lam )$ stands for the subset of $\cC (\Lam )$ formed by $\bx\subset\Lam$ with
$$\min\;\Big[|x -x'|:\;x,x'\in\bx ,\;x\neq x' \Big]\geq\td;\eqno (1.1.25)$$
cf. (1.1.20). Here and later on, the symbol $\sharp$ is used for the cardinality of a given set.
The same meaning is attributed to the notation $\cC_\td(\bbR^d)$ and $\cC_\td(\Lamc)$
(here we bear in mind finite or countable sets $\bx\subset\bbR^d$ and $\bx'\subset\Lamc$, 
respectively, obeying (1.1.25)). Therefore,  condition (1.1.20) is equivalent to
writing $\bx (\Lamc)\in\cC_\td (\Lamc)$. Points $\bx$, $\bx'$ are called, as before, classical
configurations (CCs).

\medskip

\medskip

{\bf 1.2. The thermodynamic limit. The shift-invariance property in two dimensions.} The 
key concept of Statistical Mechanics is the thermodynamic limit; in the context of this work 
it is $\lim\limits_{\Lam\nearrow\bbR^d}$.  The quantities and objects established as limiting 
points in the course of this limit are often referred to as infinite-volume ones (e.g., an infinite-volume 
RDM or GS). The existence and uniqueness of a limiting object is often interpreted as absence 
of a phase transition. On the other hand, a multitude of such objects (viz., depending on the 
boundary conditions for the Hamiltonian or the choice of external CCs) is treated 
as a sign of a phase transition. 

However, there exists an elegant alternative where infinite-volume objects are identified 
in terms which, at least formally, do not invoke the thermodynamic limit. For classical 
systems, this is the DLR equations and for  so-called quantum spin systems -- the KMS 
construction. (The latter involves an infinite-volume dynamics, a concept that is not affected by 
absence or presence of phase transitions.)
Unfortunately, the KMS construction is not directly available for the class of quantum systems 
under consideration, since the Hamiltonians $H_{n,\Lam}$ and 
$H_{n,\Lam |\bx (\Lamc )}$ are not bounded.

In this paper we propose a construction generalising the classical DLR equation (see Section 2.3).
A justification of this construction will be given in Ref. \cite{SKS} where we  
establish the shift-invariance property for the emerging objects (the RDMs and GSs) in dimension two
(i.e., for $d=2$).\footnote{In the one-dimensional case ($d=1$) our construction yields
uniqueness of an infinite-volume RDM and GS. This follows from earlier results;
cf. \cite{Su1} and \cite{Su2}.} For reader's convenience, we state the results from \cite{SKS} below,
after we give the assertions of the present paper.

The symbol $\Box$ marks the end of an assertion. The first result claimed in this work is

\medskip

\medskip

{\bf Theorem 1.1.} {\sl  Assume inequality {\rm{(1.1.19)}}. $\forall$ cube $\Lam_0$
(see Eqn {\rm{(1.1.16)}}), the 
family of RDMs $\{R^{\Lam_0}_{\Lam |\bx (\Lamc )},\Lam\nearrow\bbR^d\}$
is compact in the trace-norm operator topology in $\cH (\Lam_0)$, for any choices of
CCs $\bx (\Lamc )\in\cC_\td(\Lamc)$ (i.e., satisfying {\rm{(1.1.20)}}).
Any limit-point operator
$R^{\Lam_0}$ for $\{R^{\Lam_0}_{\Lam |\bx (\Lamc )}\}$ is a positive-definite operator in
$\cH (\Lam_0)$ of trace $1$. Furthermore, let $\Lam_1\subset\Lam_0$ be a pair of cubes 
and $R^{\Lam_1}$, $R^{\Lam_0}$ be a pair of limit-point RDMs  such that 
$$R^{\Lam_1}=\lim\limits_{l\to +\infty}R^{\Lam_1}_{\Lam (l)|\bx (\Lam (k)^{\rm c}}
\hbox{ and }R^{\Lam_0}=\lim\limits_{l\to +\infty}R^{\Lam_0}_{\Lam (l)|\bx (\Lam (k)^{\rm c}}
\eqno (1.2.1)$$
for a sequence of cubes $\Lam (l)=[-L(l),L(l)]^{\times d}$ where $l=1,2,\ldots$, $L(l)\nearrow\infty$ 
and external CCs $\bx (\Lam (l)^{\rm c})\in\cC_\td(\Lam (l)^{\rm c})$. 
Then  $R^{\Lam_1}$ and $R^{\Lam_0}$ satisfy the compatibility property 
$$R^{\Lam_1}=\tr_{\cH (\Lam_0\setminus\Lam_1)}R^{\Lam_0}. \eqno (1.2.2)$$}
$\Box$

\medskip

\medskip

In future, referring to external CCs $\bx (\Lam^{\rm c})$ and $\bx (\Lam (l)^{\rm c})$,
we always assume that $\bx (\Lam^{\rm c})\in\cC_\td(\Lamc)$ and 
$\bx (\Lam (l)^{\rm c})\in\cC_\td(\Lam (l)^{\rm c})$, that is, the condition (1.1.20) is satisfied.

\medskip

The subsequent sections of the paper carry the proof of Theorem 1.1.

\medskip

A direct corollary of Theorem 1.1 is the construction of a limit-point Gibbs state $\varphi$.
To this end, it suffices to consider a countable family of cubes $\Lam_0(l_0)=
\left[-L_0l_0,L_0l_0\right]^{\times d}$ of side-length $2L_0l_0$, where $L_0\in (0,\infty )$ 
is fixed and $l_0=1,2,\ldots$, 
centered at the origin. By invoking a diagonal process, we can guarantee that,
given a family of external CCs  $\bx (\Lamc )$, one can extract a sequence 
$\Lam (l)\nearrow\bbR^d$ such that (i) $\forall$ positive integer $l_0$ $\;\exists\;$ 
the trace-norm limit 
$$R^{\Lam_0(l_0)}=\lim\limits_{l\to +\infty}
R^{\Lam_0(l_0)}_{\Lam (l)|\bx (\Lam (l)^{\rm c})}.\eqno (1.2.3)$$ 
and (ii) for the limiting operators relation (1.2.1) holds true with 
$\Lam_1=\Lam_0(l_1)$ and $\Lam_0=\Lam_0(l_0)$
whenever $l_1<l_0$. This enables us to define an infinite-volume Gibbs state 
$\varphi$ by setting 
$$\varphi (A)=\lim\limits_{l\to\infty}\varphi_{\Lam (l)}(A)=
{\tr}_{\cH (\Lam_0(l_0))}\big(AR^{\Lam_0(l_0)}\big),\;\;A\in\fB (\Lam_0(l_0)).\eqno (1.2.4)$$
More precisely, $\varphi$ is a state of the quasilocal C$^*$-algebra $\fB (\bbR^d)$ defined as the 
norm-closure of the inductive limit $\fB^0(\bbR^d)$:
$$\fB=\left( \fB^0(\bbR^d)\right)^-,\;\; \fB^0(\bbR^d)=
\operatornamewithlimits{\hbox{ind lim}}\limits_{\Lam\nearrow\bbR^d}
\fB (\Lam ).\eqno (1.2.5)$$
Moreover, $\phi$ is determined by a family of finite-volume RDMs $R^{\Lam_0}$
acting in $\cH(\Lam_0)$ where $\Lam_0\subset\bbR^d$ is an arbitrary cube of the form (1.1.16) 
and satisfying the compatibility property (1.2.2). 

As was said, in paper \cite{SKS} we establish the property of shift-invariance of the
limit-point Gibbs states $\varphi$ when the dimension $d=2$. Observe that $\forall$ 
cube $\Lam_0$ as in (1.1.16) and vector $s=\left({\ts}^1,\ldots ,{\ts}^d\right)\in\bbR^d$, 
the Fock spaces $\cH (\Lam_0)$ and $\cH ({\tS}(s)\Lam_0)$ are related through a pair 
of mutually inverse shift isomorphisms of Fock spaces
$${\tU}^{\Lam_0}(s):\;\cH (\Lam_0)\to\cH ({\tS}(s)\Lam_0)\hbox{ and }
{\tU}^{\tS(s)\Lam_0}(-s):\; \cH ({\tS}(s)\Lam_0)\to\cH (\Lam_0).$$
Here ${\tS}({\ts})$ stands for the shift isometry $\bbR^d\to\bbR^d$:
$${\tS}(s):\;y\mapsto y+s,\;\;y\in\bbR^d,\eqno (1.2.6)$$ 
and ${\tS}(s)\Lam_0$ for the image of $\Lam_0$:
$$\begin{array}{l}{\tS}(s)\Lam_0
=\left[-L_0+{\tc}^1_0+{\ts}^1,{\ts}^1+{\tc}^1_0+L^0\right]\\
\quad\qquad\qquad\times\cdots\times
\left[-L_0+{\tc}^d_0+{\ts}^d,{\ts}^d+{\tc}^d_0+L^0\right].\end{array} \eqno (1.2.7)$$
The isomorphisms ${\tU}^{\Lam_0}(s)$ and 
${\tU}^{\tS(s)\Lam_0}(-s)$ are determined by
$$\begin{array}{c}\left({\tU}^{\Lam_0}(s)\phi_n\right) (\ux_1^n)
=\phi_n ({\tS}(-s)\ux_1^n),\;\;\ux_1^n\in\left(\tS(s)\Lam_0\right)^n,\;\;\phi_n\in 
{\rm L}_2^{\rm{sym}}\left((\Lam_0)^n_\td\right),\\ \;\\
\left({\tU}^{\tS(s)\Lam_0}(-s)\phi_n\right) (\ux_1^n)=
\phi_n ({\tS}(s)\ux_1^n),
\ux_1^n\in\left(\Lam_0\right)^n,\phi_n\in 
{\rm L}_2^{\rm{sym}}\left((\tS(s)\Lam_0)^n_\td\right),\\ \;\end{array}\eqno (1.2.8)$$
where $n=0,1,\ldots $.

\medskip

In two dimensions,  the main result of \cite{SKS} is 

\medskip

\medskip

{\bf Theorem 1.2.} {\sl Suppose that $d=2$ and the condition {\rm{(1.1.19)}} is fulfilled. Assume in 
addition that the potential $V$ satisfies  
$${\ov V}^{\,(2)}=\max\,\big[\left| V''(r)\right|: \td\leq r\leq\tR\big]
.\eqno (1.2.9)$$
Then any limit-point Gibbs state $\varphi$ is shift-invariant: $\forall$ $s=(\ts^1,\ts^2)\in\bbR^{2}$
$$\varphi (A)=\varphi ({\tS}(s)A),\;\;A\in\fB(\bbR^{2}).\eqno (1.2.10)$$
Here ${\tS}(s)A)$ stands for the shift of the argument $A$: if $A\in\fB (\Lam_0)$ where $\Lam_0$
is a square $[-L_0+\tc_1,\tc_1+L_0]\times [-L_0+\tc_2,\tc_2+L_0]$ then
$${\tS}(s)A ={\tU}^{\tS(s)\Lam_0}(-s)A\,{\tU}^{\Lam_0}(s)\in\fB({\tS}(s)\Lam_0).$$ 
In terms of the RDMs $R^{\Lam_0}$:
$$R^{\tS(s)\Lam_0} ={\tU}^{\Lam_0}(s)R^{\Lam_0}\,{\tU}^{\tS(s)\Lam_0}(-s).\eqno (1.2.11)$$}
$\Box$ 

\medskip 

\medskip

{\bf Remark.} Theorems 1.1 and 1.2  can be extended to the case of systems with several
particle types. This line of study is pursued in \cite{SS} where the case of non-negative interaction
potentials is considered, including a `pure hard-core` two-body repulsion. A notable example is 
a quantum (bosonic) version of a Widom--Rowlinson model; cf. \cite{CP}, \cite{I}. This model exhibits
a `spatial` phase transition which is expected to fit the theory developed in the present paper.
On the other hand, an important question that remains open is how the present theory (more precisely, its 
eventual extension) can explain 
the phenomenon of Bose--Einstein condensation. 

\medskip

\medskip

{\bf 1.3. The integral kernels of Gibbs' operators and RDMs.} Let us return to a general
value of dimension $d$. We will assume condition (1.1.19) without stressing this every time again.
According to the adopted realization of the Fock space $\cH (\Lam )$ as
${\rm L}_2(\cC_\td(\Lam ))$ (see (1.1.24)), its elements are 
represented by functions 
$\phi_\Lam :\;\bx(\Lam )\in\cC_\td(\Lam )\mapsto\phi_\Lam (\bx (\Lam ))\in{\bbC}$,
with 
$$\diy\int_{\cC (\Lam )}
\left|\phi_\Lam (\bx (\Lam ) )\right|^2{\rd}\bx (\Lam )<\infty .
\eqno (1.3.1)$$
The space
$\cH (\Lam_0)$ is described in a similar manner: here we will use  a short-hand
notation $\bx_0$ and $\by_0$ instead of  $\bx (\Lam_0),\by (\Lam_0)\in\cC_\td(\Lam_0)$. 
 
The  first step in the proof of Theorems 1.1 is to reduce its assertions to statements 
about  the integral kernels $F^{\Lam_0}_\Lam$, $F^{\Lam_0}_{\Lam |\bx (\Lamc )}$ and 
$F^{\Lam_0}$ which define the RDMs $R^{\Lam_0}_\Lam$, $R^{\Lam_0}_{\Lam |\bx (\Lamc )}$
and their infinite-volume counterpart $R^{\Lam_0}$; we call these kernels RDMKs for short. Indeed, 
$R^{\Lam_0}_\Lam$, $R^{\Lam_0}_{\Lam |\bx (\Lamc )}$ and $R^{\Lam_0}$ are integral 
operators:
$$\left(R^{\Lam_0}_\Lam\phi_\Lam\right) (\bx_0)=\int_{{\cC}_\td(\Lam )}
F^{\Lam_0}_\Lam (\bx_0,\by_0)
\phi_\Lam (\by_0){\rd}\by_0,\eqno (1.3.2)$$
$$\left(R^{\Lam_0}_{\Lam |\bx (\Lamc )}\phi_\Lam\right) (\bx_0)
=\int_{{\cC}_\td(\Lam )} F^{\Lam_0}_{\Lam |\bx (\Lamc )}(\bx_0,\by_0)
\phi_\Lam (\by_0){\rd}\by_0\eqno (1.3.3)$$
and
$$\left(R^{\Lam_0}\phi_\Lam\right) (\bx_0)=\int_{{\cC}_\td(\Lam )}
F^{\Lam_0}(\bx_0,\by_0)
\phi_\Lam (\by_0){\rd}\by_0.\eqno (1.3.4)$$
The RDMKs $F^{\Lam_0}_\Lam (\bx_0,\by_0)$ and $F^{\Lam_0}_{\Lam |\bx (\Lamc )}(\bx_0,\by_0)$
-- and ultimately $F^{\Lam_0}(\bx_0,\by_0)$ -- admit an FK representation
providing a basis for the future analysis. Here we state properties of these kernels
in Theorems 1.3 where we adopt a setting from Theorem 1.1:

\medskip

\medskip

{\bf Theorem 1.3.} {\sl $\forall$  pair of cubes $\Lam_0\subset\Lam$ and 
CCs $\bx (\Lamc )\in\cC_\td(\Lamc)$,  
the family of RDMKs $F^{\Lam_0}_{\Lam |\bx (\Lamc )}(\bx_0,\by_0)$
is compact in the space of continuous functions $C^0\left({\cC}_\td(\Lam_0)\times{\cC}_\td
(\Lam_0)\right)$. Any limit-point function 
$$(\bx_0,\by_0)\in{\cC}_\td(\Lam_0)\times{\cC}_\td(\Lam_0)\mapsto 
F^{\Lam_0}(\bx_0,\by_0)\eqno (1.3.5)$$
determines a positive-definite operator $R^{\Lam_0}$  in $\cH (\Lam_0)$ of trace $1$ 
(a limit-point RDM). Furthermore, let $\Lam_1\subset\Lam_0$ be a pair of cubes 
and $F^{\Lam_1}$, $F^{\Lam_0}$ a pair of limit-point RDMKs  such that 
$$F^{\Lam_1}=\lim\limits_{l\to +\infty}F^{\Lam_1}_{\Lam (l)}
\hbox{ and }F^{\Lam_0}=\lim\limits_{l\to +\infty}F^{\Lam_0}_{\Lam (l)|\bx (\Lam (l)^{\rm c})}
\eqno (1.3.6)$$
in $C^0\left({\cC}_\td(\Lam_0)\times{\cC}_\td(\Lam_0)\right)$ for a sequence of cubes 
$\Lam (l)\nearrow\bbR^d$ and 
external CCs $\bx\left(\Lam (l)^{\rm c}\right)$. Then the corresponding 
limit-point RDMs  $R^{\Lam_1}$ and $R^{\Lam_0}$ obey {\rm{(1.2.2)}}.}  $\qquad\Box$ 

\medskip

\medskip

Theorem 1.3 implies Theorem 1.1 with the help of Lemma 1.5 from \cite{KS1} (going 
back to Lemma 1 in \cite{Su1}). 
Therefore we focus on the proof of Theorems 1.3. In fact, we will establish
the properties for more general objects -- FK-DLR functionals.
The rest of the paper is organized as follows.  In Section 2.1 we introduce the    
 FK-representation for the RDMKs   $F^{\Lam_0}_\Lam (\bx_0,\by_0)$
 and $F^{\Lam_0}_{\Lam |\bx (\Lamc )}(\bx_0,\by_0)$ and in Section 2.2 the
 FK-representation for their infinite-volume counterparts   
$F^{\Lam_0}(\bx_0,\by_0)$. On the basis of this representation we
define the class of FK-DLR states (more generally, FK-DLR functionals)
and state Theorems 2.1 and 2.2 extending the assertions of Theorems 1.4 and 1.5 
to this class. Section 3 contains the proofs.
\vskip 1 truecm

\section{The FK representation and the FK-DLR equation}

\medskip

\medskip

{\bf 2.1. The background of the FK-representation.} The symbol  $\triangle$ marks below the 
end of a definition. We begin with three definitions
(see Definitions 2.1.1--2.1.3). In part, these definitions repeat pieces of papers \cite{KS1},
\cite{KS2}, \cite{KSY2}.
\medskip

{\bf Definition 2.1.1.} ({\it Path spaces}.) As above, $x,y$ stands 
for points in $\bbR^d$, $\ux =\ux_1^n=\{x(1),\ldots ,x(n)\} $ and $\uy 
=\uy_1^n=\{y(1),\ldots ,y(n)\} $ for points in $\Lam^n$. Next, $\gam =\gam_n$ denotes
a permutation of the $n$th order, $\gam\uy =\{y(\gam (1)),\ldots ,$\\ $y(\gam (n))\}$ 
stands for the vector with permuted entries and  $\bx (\Lam )$ for a
point in ${\cC}(\Lam )$ (i.e., a finite subset of $\Lam$).  Furthermore, we will use the 
following system of notation:

(i)\;\; $\ocW^{\,k\beta}(x,y)$ -- the space of continuous paths $\oom=\oom_{x,y}:\;[0,k\beta ]\to\bbR^d$
of time-length $k\beta$  (the parameter $k$ is called the time-length multiplicity), with 
$\oom^*(0)=x$, $\oom^*(k\beta )=y$, where $k=1,2,\ldots$;

(ii)\;\; $\ocW^{\,*}(x,y)=\operatornamewithlimits{\cup}\limits_{k\geq 1}\ocW^{k\beta}(x,y)$ -- 
the space of continuous paths $\oom^*=\oom^*_{x,y}:\;[0,\beta ]\to\bbR^d$
of a variable time-length $k\beta$, with $\oom^*(0)=x$, $\oom^*(k\beta )=y$;

(iii)\;\; $\cW^*(x)=\ocW^{\,*}(x,x)$ -- the space of loops (closed paths) $\om^*=\om^*_x$
with $\om^*(0)=\om^*(\beta )=x$;

(iv)\;\; $\ocW^*(\ux,\uy)=\operatornamewithlimits{\times}\limits_{1\leq j\leq n}\ocW^*(x(j),y(j))$ --
the space of (ordered) path collections $\oOm^*=\{\oom^*(1),\ldots ,\oom^*(n)\}$ where
$\oom (j)\in\ocW^*(x(j),y(j))$;

(v)\;\; $\ucW^*(\ux,\uy)=\operatornamewithlimits{\cup}\limits_{\gam_n}\ocW^*(\ux,\gam_n
\uy)$ -- the space of path collections $\oOm^*$ with permuted
endpoints;

(vi)\;\; $\cW^*(\bx )=\operatornamewithlimits{\times}\limits_{x\in\bx}\cW^*(x)$ --
the space of loop collections $\Om^*(\bx )$\\ $=$ $\{\om^*(x), x\in\bx\}$ 
($\Om^*$ for short) with a given (finite) initial/end-point
CC $\bx \in{\cC}_\td(\bbR^d )$, where $\om (x)\in \cW^*(x)$;

(vii)\;\; $\cW^*(\Lam )=\operatornamewithlimits{\cup}\limits_{\bx\in{\cC}_\td(\Lam )}\cW^*(\bx )$ --
the space of loop collections $\Om^*=\Om^*(\Lam )$\\ $=$ $\{\om^*(x),\;x\in\bx\}$  
with various  initial/end-point CCs $\bx =\bx (\Lam )\in{\cC}_\td(\Lam )$. Sometimes it 
will be helpful to stress that an element $\Om^*\in\cW^*(\Lam )$ is a pair 
$\big[\bx (\Lam ),\Om^*(\bx (\Lam ))\big]$ where $\Om^*(\bx (\Lam ))\in\cW^*(\bx (\Lam ))$ 
and treat a loop  
$\om^*(x)\in \cW^*_x$ (or rather its shift ${\tS}(-x)\om^*(x)\in \cW^*(0)$) as a `mark'  for point 
$x\in\bx (\Lam )$. (Here and below, the loop ${\tS}(s)\om^*$ is defined by $\left({\tS}(s)\om^*\right)(\ttt )=
\om^*(\ttt )+s$, $s\in\bbR^d$, $\ttt\in [0,\beta k(\om^*)]$.) Such a view is useful when we work with 
probability measures (PMs) on $\cW^*(\Lam )$: in the probabilistic terminology these PMs represent two-dimensional random marked point processes (RMPPs) in $\Lam$ with marks from $\cW^*(0)$, the space of loops starting and finishing at $0$.

An element $\oOm^*$ from $\ocW^{\,*}(\ux,\uy)$ is called a path configuration (PC), with the initial/terminal 
CCs $\ux$, $\uy$.
Likewise, an element $\Om^*\in \cW^* (\Lam )$ is called a loop configuration (LC) over $\Lam$; if 
$\Om^*\in \cW^* (\bx(\Lam ))$, we say that $\bx (\Lam)$ is the initial CC for $\Om^*$.
The time-length multiplicity of  a path $\oom^*\in\ocW_{x,y}$ is denoted by $k(\oom^*)$.
The next series of definitions is introduced for a fixed $\ttt\in [0,\beta ]$. Namely,
given  a path $\oom^*\in\ocW_{x,y}$, we call the set 
$$\{\oom^*({\ov l}\beta +\ttt ),
\;{\ov l}=0,\ldots ,k(\oom^*)-1\}\subset\bbR^d$$ 
the $\ttt$-section of  $\oom^*$ and denote it by 
$\{\oom^*\}(\ttt )$. Next, given a PC  $\oOm^*=\{\om^*(1),\ldots ,\om^*(n)\}\in \ocW^*_{\ux,\uy}$, 
the $\ttt$-section for $\oOm^*$ is defined as the union 
$$\{\oOm^*\}(\ttt )=
\operatornamewithlimits{\cup}\limits_{1\leq j\leq n}\oom^*(j,\ttt)$$ 
where $\{\oom^*(j)\}(\ttt )= 
\{\oom^*(j,{\ov l}(j)\beta +\ttt ),\;0\leq{\ov l}(j)<k(\oom^*(j))\}$ is the $\ttt$-section for path $\oom (j)\in\oOm^*$ 
(thus, $\{\oOm^*\}(\ttt )$ again is a subset of $\bbR^d$). Likewise, given 
a LC $\Om^*=\{\om^*(x), x\in\bx (\Lam )\}\in \cW^*(\bx(\Lam ))$, the set 
$$\{\Om^*\}(\ttt )=\operatornamewithlimits{\cup}\limits_{1\leq j\leq n}\{\om^*(x)\}(\ttt )\subset\bbR^d$$ 
is called
the $\ttt$-section of $\Om^*$. Here $\{\om^*(x)\}(\ttt )=\{\om^*(x,l\beta +\ttt ),\;0\leq l<k(\om^*(x)),\;x\in
\bx (\Lam )\}\subset\bbR^d$  is the $\ttt$-section of loop $\om^*(x)$. The concept of a $\ttt$-section
plays a subsidiary r\^ole in this paper but becomes essential in \cite{SKS}. 

Similar definitions and terms will be used for a cube $\Lam_0\subset\Lam$ or the 
set-theoretical difference $\Lam\setminus\Lam_0$. 

All path/loop spaces $\ocW^{\,k\beta}(x,y)$, $\ocW^{\,*}(x,y)$, $\cW^*(x)$, $\ocW^*(\ux,\uy)$, $\ucW^*(\ux,\uy)$, $\cW^*(\bx (\Lam ))$, $\cW^*(\Lam )$ from (i)--(vii) contain subsets  $\ocW^{\,k\beta}_\td(x,y)$, $\ocW^{\,*}_\td(x,y)$, $\cW^*_\td(x)$, $\ocW^*_\td(\ux,\uy)$, $\ucW^*_\td(\ux,\uy)$, $\cW^*_\td(\bx (\Lam ))$ and $\cW^*_\td(\Lam )
=\operatornamewithlimits{\cup}\limits_{\bx (\Lam )\in{\cC}_\td(\Lam )}\cW^*_\td(\bx (\Lam ))$ extracted by
the condition that $\forall$ $\ttt\in [0,\beta ]$ no two distinct points in the $t$-section lie at a Euclidean distance
$\leq a$.  In other words, all sections $\{\oom^*\}(\ttt )$, $\{\oOm^*\}(\ttt )$ $\{\Om^*\}(\ttt )$ are (finite) CCs lying in $\cC_\td(\bbR^d)$. $\qquad\triangle$

\medskip

{\bf Definition 2.1.2.} ({\it Path measures}.) The spaces introduced in Definition 2.1.1 are
equipped with standard sigma-algebras (generated by cylinder subsets and operations
on them). We consider various measures on these sigma-algebras:

(i)\;\; ${\ov{\bbP}}^{\,k\beta}_{x,y}$ -- the (non-normalized)  measure on $\ocW^{k\beta}_{x,y}$ (the Wiener
bridge of time-length $k\beta$), with ${\ov{\bbP}}^{\,k\beta}_{x,y}\left(\ocW^{\,k\beta}_{x,y}\right)$ $=$
$(2\pi k\beta )^{-1}\exp\,\Big[ -|x-y|^2\big/(2k\beta)\Big]$;
 
(ii)\;\; ${\ov{\bbP}}^{\,*}_{x,y}$ -- the sum-measure $\sum\limits_{k\geq 1}{\ov{\bbP}}^{\,k\beta}_{x,y}$  on
$\ocW^{\,*}_{x,y}$;

(iii)\;\; ${\bbP}^*_x={\ov{\bbP}}^{\,*}_{x,x}$ -- the sum-measure $\sum\limits_{k\geq 1}{\ov{\bbP}}^{\,k\beta}_{x}$  on $\cW^*_{x}$;

(iv)\;\; ${\ov{\bbP}}^{\,*}_{\ux,\uy}=\operatornamewithlimits{\times}\limits_{1\leq j\leq n}{\ov{\bbP}}^*_{x(j),y(j)}$ --
the product-measure on $\ocW^{\,*}_{\ux,\uy}$ (a vector Wiener bridge) under which the components 
$\om^*(j)$ are independent;

(v)\;\; ${\ubbP}^*_{\ux,\uy}=\sum\limits_{\gam_n}{\ov{\bbP}}^*_{\ux,\gam_n\uy}$ -- the sum-measure on 
${\ucW}^*_{\ux,\uy}$;

(vi)\;\; ${\bbP}^*_{\bx}=\operatornamewithlimits{\times}\limits_{x\in\bx}{\bbP}^*_x$ --
the product-measure on $\cW^*(\bx)$;

(vii)\;\; ${\rd}\Om^*(\Lam )={\rd}\bx(\Lam )\times{\bbP}^*_{\bx (\Lam )}({\rd}\Om^*)$ -- the measure on 
$\cW^*(\Lam )$ where\\ ${\rd}\bx(\Lam )$ is the Lebesgue--Poisson measure on ${\cC}(\Lam )$
(cf. (1.3.2)). Sometimes we will write ${\rd}^\Lam\bx(\Lam )$ and ${\rd}^\Lam\Om^*(\Lam )$ in order
to stress the dependence upon $\Lam$ (in particular, the notation ${\rd}^{\Lam_0}\Om^*_0$ will
be used); other types of sets (the difference of two cubes) will also be
employed. $\qquad\triangle$ 

\medskip

\medskip

{\bf Definition 2.1.3.} ({\it Energy-related functionals}.) Given a path $\oom^*\in\ocW^{\,*}_\td(x,y)$, 
the functional $h (\oom^*)$ is defined by
$$\begin{array}{cl}h(\oom^*)&=\diy\int_0^\beta\rd\ttt \sum_{0\leq l<l'<k(\oom^*)}V\left(\left|\oom^*(t+l\beta )-
\oom^*(t+l'\beta )\right|\right)\\ \;&= \diy\int_0^\beta\rd\ttt E\big(\{\oom^*\}(\ttt )\big).\end{array}\eqno (2.1.1)$$
Here, for a given finite CC $\bz\in\cC_\td(\bbR^d)$, we set:
$$E(\bz)=\frac{1}{2}\sum_{(z,z')\in\bz\times\bz}V(|z-z'|).\eqno (2.1.2)$$
The quantity $h(\oom )$ can be interpreted as an energy of path $\oom$.

The energy of interaction between two paths, $\oom^*\in\ocW^{\,*}_\td(x,y)$ and 
${\oom^*}'\in\ocW^{\,*}_\td(x',y')$, is determined by
$$\begin{array}{cl}h(\oom^*,{\oom^*}' )&=\diy\int_0^\beta\rd\ttt \sum_{0\leq l<k(\oom^*)}
\sum_{0\leq l'<k({\oom^*}' )}V\left(\left|\oom^*(t+l\beta )-
{\oom^*}' (t+l'\beta )\right|\right)\\ \;&
=\diy\int_0^\beta\rd\ttt  E\Big(\{\oom^*\}(\ttt )\Big\|\{{\oom^*}'\} (\ttt )\Big).\end{array}\eqno (2.1.3)$$
Here, for given pair of CCs $\bz,\bz' \in\cC_\td(\bbR^d)$, such that 
$\bz\cup\bz'\in\cC_\td(\bbR^d)$, $\bz\cap\bz'=\emptyset$ and at least one of them 
is finite, we set:
$$E(\bz ||\bz' )=\sum_{(z,z' )\in\bz\times\bz'}V(|z-z' |).\eqno (2.1.4)$$
The definitions (2.1.1) and (2.1.3) holds for loops as well, obviously.

Next, for a PC $\oOm^*=\{\oom^*(1),\ldots ,\oom^*(n)\}\in \ocW^{\,*}_\td(\ux,\uy )$
and a LC $\Om^*=\{\om^*(x)\}\in \cW^*_\td(\bx (\Lam ))$, the energy 
$h( \oOm^*)$ of $\oOm^*$ and the energy $h( \Om^*)$ of $\Om^*$ are  defined as
$$h(\oOm^*)=\sum_{1\leq j\leq n}h(\oom^*(j))+\sum_{1\leq j<j' \leq n}h(\oom^*(j),\oom^*(j' ))
\eqno (2.1.5)$$
and
$$h(\Om^*)=\sum_{x\in\bx (\Lam )}h(\om^*(x))+\frac{1}{2}\sum_{x,x'\in\bx (\Lam ):\;x\neq x'}
h(\oom^*(x),\oom^*(x' )).\eqno (2.1.6)$$
We will also need the energy for various combined collections of PCs, LCs and CCs. Viz., 
for $\oOm^*\in \ocW^*_\td(\ux,\uy )$ where $\ux,\uy\in\Lam^n$ and 
$\Om^*=\{\om^*(x)\}\in \cW^*_\td(\Lam )$, 
$$h(\oOm^*\vee\Om^*)=h( \oOm^*)+h( \Om^*)+h( \oOm^*||\Om^*)\eqno (2.1.7)$$ 
where
$$h( \oOm^*||\Om^*) =\int_0^\beta\rd\ttt E(\oOm^*(\ttt )||\Om^*(\ttt )).\eqno (2.1.8)$$
Finally, for  $\bx (\Lamc )\in\cC_\td(\Lamc )$, 
$$h(\oOm^{\,*}\vee\Om^*|\bx (\Lamc ))=h( \oOm^*\vee \Om^*)+h( \oOm^*\vee\Om^*||\bx (\Lamc ))
\eqno (2.1.9)$$
where
$$h( \oOm^{\,*}\vee\Om^*||\bx (\Lamc ))=\int_0^\beta\rd\ttt E(\{\oOm^*\}(\ttt )\cup\{\Om^*\}(\ttt )||\bx(\Lamc )). 
\eqno (2.1.10)$$

Finally, we introduce the functionals $K$, $L$ and $\alpha_\Lam$, for path and LCs :
$$K(\oOm^*)=\sum_{\oom^*\in\oOm^*} k(\oom^*),\;\;K(\Om^*)=\sum_{\om^*\in\Om^*} k(\om^*),
\eqno (2.1.11)$$
and
$$L(\Om^*)=\prod\limits_{\om^*\in\Om^*} k(\om^*).
\eqno (2.1.12)$$
The presence of Dirichlet's boundary conditions is manifested in the indicators 
$$\alpha_\Lam  (\oOm^*)=\prod\limits_{\oom^*\in\oOm^*}\alpha_\Lam  (\oom^*),
\;\;\alpha_\Lam  (\Om^*)=\prod\limits_{\om^*\in\Om^*}\alpha_\Lam  (\om^*),\eqno (2.1.13)$$
where
$$\alpha_\Lam  (\oom^*)={\mathbf 1}\Big(\oom^*(\ttt )\in\Lam\;\;\forall\;\ttt\in\big[0,k(\oom^*)
\beta\big] \Big).\eqno (2.1.14)$$
$\triangle$
 
\medskip

\medskip

{\bf 2.2. The FK-representation in a cube.}
As follows from well-known results about the operator $H_\Lam$ (see, e.g., \cite{Gi},
\cite{S1}), we have the following properties listed in Lemmas 2.2.1 and 2.2.2. 

\medskip

\medskip

{\bf Lemma 2.2.1.} {\sl For a chosen 
external CC $\bx (\Lamc )$ defining the self-adjoint operators 
$H_{n,\Lam |\bx (\Lamc )}$, the partition function $\Xi\big[\Lam |\bx (\Lamc )\big]$
(see {\rm{(1.1.23))}} admits the following representation:
$$\begin{array}{r}\diy\Xi\big[\Lam |\bx (\Lamc )\big]=\int_{\cW^*_\td(\Lam)}{\rd}
\Om^*_\Lam\alpha_\Lam (\Om^*_\Lam )\qquad\qquad\qquad\qquad{}\\
\diy\times\frac{z^{K(\Om^*_\Lam)}}{L(\Om^*_\Lam)}\exp\;\left[
-h\Big(\Om^*_\Lam\big|\bx (\Lamc )\Big)\right]\,.\end{array}\eqno (2.2.1)$$
See Definitions {\rm{2.1.1(iii),(vi),(vii)}} and  {\rm{2.1.2(iii),(vi),(vii)}}.

Moreover, for the corresponding RDMK $F^{\Lam_0}_{\Lam |\bx (\Lamc )}$ (see {\rm{(1.3.3),
(1.3.4)}}) we have that for $\bx_0,\by_0\in\cC_\td(\Lam_0)$ with $\sharp\;\bx_0=\sharp\;\by_0$:
$$\begin{array}{l}\diy F^{\Lam_0}_{\Lam |\bx (\Lamc )}(\bx_0,\by_0)\\
\quad\diy =
\int_{\ucW^*_{\,\td}(\bx_0,\by_0)}
{\ubbP}^*_{\;\bx_0,\by_0}({\rd}\uOm^*_0)\chi^{\Lam_0}(\uOm^*_0)
\alpha_\Lam (\uOm^{\,*}_0)z^{K(\uOm^*_0)}
{\wh q}^{\Lam_0}_{\Lam |\bx (\Lamc )}(\uOm^*_0).\end{array}\eqno (2.2.2)$$
Cf. Definitions {\rm{2.1.1(v)}} and {\rm{2.1.2(v)}}. Here
$${\wh q}^{\Lam_0}_{\Lam |\bx (\Lamc )}(\uOm^*_0)=
\frac{{\wh\Xi}^{\Lam_0,\uOm^*_0}\big[\Lam\setminus\Lam_0 |\bx (\Lamc )\big]
}{\Xi\big[\Lam |\bx (\Lamc )\big]}\,,\;\uOm^*_0\in {\ucW^{\,*}_{\,\td}(\bx_0,\by_0)},\eqno (2.2.3) $$ 
$\Xi\big[\Lam |\bx (\Lamc )\big]$ is defined as in {\rm{(2.2.1)}} and 
$$\begin{array}{l}
\diy{\wh\Xi}^{\Lam_0,\uOm^*_0}\big[\Lam\setminus\Lam_0 |\bx (\Lamc )\big]\\
\quad\diy =\int_{\cW^*_\td(\Lam )}{\rd}\Om^*_{\Lam\setminus\Lam_0}
{\mathbf 1}\Big(\Om^*_{\Lam\setminus\Lam_0}\in\cW^*_\td(\Lam\setminus\Lam_0)\Big)
\chi^{\Lam_0}(\Om^*_{\Lam\setminus\Lam_0})\\
\qquad\qquad\diy\times\alpha_\Lam (\Om^*_{\Lam\setminus\Lam_0})
\frac{z^{K(\Om^*_{\Lam\setminus\Lam_0})}}{L(\Om^*_{\Lam\setminus\Lam_0})}\exp\;\left[
-h\Big(\uOm^*_{\,0}\vee\Om^*_{\Lam\setminus\Lam_0}\big|\bx (\Lamc )\Big)\right].
\end{array}\eqno (2.2.4)$$
Functionals $K$ and $L$ are as in {\rm{(2.1.11), (2.1.12)}}.
Next, $\chi^{\Lam_0}(\uOm^*_0\vee\Om^*_{\Lam\setminus\Lam_0})$ stands for the indicator requiring that no 
path $\oom^*$ or loop $\om^*$ from the whole collection 
enters the cube $\Lam_0$ at `control' time points $l\beta$ with $1\leq l<k$, where
$k$ equals $k(\oom^*)$ or $k(\om^*)$.
 
Namely,  for a PC
$\uOm^*_0=\{\oom^*(1),\ldots ,\oom^*(n)\}\in\ucW^{\,*}(\bx_0,\by_0)$ where $\ux=\{x(1),\ldots ,x(n)\}$,
$\uy=\{y(1),\ldots ,y(n)\}\in\cC (\Lam_0)$ and a LC $\Om^*_{\Lam\setminus\Lam_0}
=\{\om^*(x),x\in\bx_{\Lam\setminus\Lam_0}\}$ 
where $\bx_{\Lam\setminus\Lam_0}\in\cC (\Lam )$,
$$\begin{array}{l}
\chi^{\Lam_0}(\uOm^*_0\vee\Om^*_{\Lam\setminus\Lam_0})\\
\;\; ={\mathbf 1}
\Big(\oom^*(j,l\beta )\in\bbR^d\setminus \Lam\;\forall\;l=1,\ldots ,k(\oom^*(j))-1,1\leq j\leq n\Big)\\
\qquad\;\times{\mathbf 1}
\Big(\om^*(x,l\beta )\in\bbR^d\setminus \Lam\;\forall\;l=1,\ldots ,k(\om^*(x))-1,x\in\bx (\Lam )\Big).
\end{array}\eqno (2.2.5)$$
Note that when $k(\oom^*(j))=1$ or $k(\oom^*(x))=1$, the above indicator yields no restriction. }
$\qquad\Box$

\medskip

\medskip

Mnemonically, the notation ${\wh\Xi}^{\Lam_0,\,\uOm^*_0}$ means the application of an indicator function
$\chi^{\Lam_0}$ in the corresponding integral, together with presence of a specific PC 
$\uOm^*_0$ in the energy functional $h\Big(\uOm^{\,*}_0\vee\Om^*\big|\bx (\Lamc )\Big)$. We can say 
that the quantity ${\wh\Xi}^{\Lam_0,\oOm^*_0}\big[\Lam\setminus\Lam_0 |\bx (\Lamc )\big]$ in (2.2.3) 
represents a restricted partition function in $\Lam\setminus\Lam_0$ in presence of a PC 
$\oOm^*_0$ and in the potential field 
generated by an external CC $\bx (\Lamc )$,  with the restriction dictated by  
$\chi^{\Lam_0}$. We would like to note that ${\wh\Xi}^{\Lam_0,\oOm^*_0}
\big[\Lam\setminus\Lam_0 |\bx (\Lamc )\big]$ is only one out of several types of partition functions 
that we will have to deal with in our analysis.  

\medskip

\medskip

{\bf Remark.} The role of the indicator functional $\chi^{\Lam_0}$ in formulas (2.2.2)--(2.2.4) is to
guarantee the condition that cube $\Lam_0$ contains exactly the right number of particles
(which is given by $\sharp\bx_0=\sharp\by_0$ and required by representations (1.3.2), (1.3.3) and --
when passing to the limit -- by (1.3.4)).

\medskip

\medskip

The aftermath of Lemma 2.1 is the emergence of a probability measure (PM), $\mu_{\Lam |\bx(\Lamc )}$, 
on the LC space $\cW^*_\td(\Lam )$ (i.e., an RMPP in $\Lam$ with marks from the loop 
space $\cW^*_\td(0)$). More precisely, $\mu_{\Lam |\bx(\Lamc )}$ is a PM on the standard (Borel)
sigma-algebra $\fW (\Lam )$ of subsets of $\cW^*(\Lam )$ supported by $\cW^*_\td(\Lam )$. 
\medskip 

{\bf Definition 2.4.} The PM $\mu_{\Lam |\bx (\Lamc )}$ is given by 
the probability density function (PDF) $f_{\Lam |\bx (\Lamc )}(\Om^*)$, $\Om^*\in \cW^*_\td(\Lam )$, where 
$$\begin{array}{cl}f_{\Lam |\bx(\Lamc )} (\Om^*)&:
=\diy\frac{\mu_{\Lam |\bx (\Lamc )}(\rd\Om^*)}{{\rd}^\Lam\Om^*}\\ 
\;&\diy\;=\alpha_\Lam (\Om^*)\;\frac{z^{K(\Om^*_\Lam)}}{L(\Om^*_\Lam)}\;\frac{\exp\;\left[
-h\Big(\Om^*_\Lam\big|\bx (\Lamc )\Big)\right]}{\Xi\big[\Lam |\bx (\Lamc )\big]},\end{array}
\eqno (2.2.6)$$
with partition function $\Xi\big[\Lam |\bx (\Lamc )\big]$ as in (1.1.21), (2.2.1). 
Furthermore, consider the restriction $\mu_{\Lam |\bx (\Lamc )}\Big|_{\fW^*_\td(\Lam_0)}$ of 
$\mu_{\Lam {}, (\Lamc )}$ to the sigma-algebra $\fW^*(\Lam_0)$ (more precisely,
to $(\cW^*_\td(\Lam_0),\fW^*(\Lam_0))$. Here $\fW^*(\Lam_0)$ is treated as a sigma-subalgebra
of $\fW^*(\Lam )$, through the map $\cW^*(\Lam )\to\cW^*(\Lam_0)$:
$$\begin{array}{l}\Om^*=\{\om^*_x,\,x\in\bx (\Lam )\subset\Lam \}
\mapsto\Om^*_{\Lam_0}=\{\om^*_x:\,x\in\bx (\Lam )\cap\Lam_0\}.
\end{array}$$ 
Then $\mu_{\Lam | \bx(\Lamc )}$ is determined by the PDF
$$\begin{array}{l}\diy f^{\Lam_0}_{\Lam |\bx (\Lamc )}(\Om^*_0)
:=\frac{\mu_{\Lam |\bx (\Lamc )}\Big|_{\cW^*(\Lam_0)}({\rd}\Om^*_0)}{{\rd}^{\Lam_0{}} \Om^*_0}\\
\diy\qquad\quad =\alpha_\Lam (\Om^*_0)\frac{z^{K(\Om^*_0)}}{L(
\Om^*_0)}
\,\frac{\Xi^{\Lam_0,\Om^*_0}\big[\Lam\setminus\Lam_0|\bx (\Lamc )
\big]}{\Xi\big[\Lam |\bx (\Lamc )\big]},\;\;\Om^*_0\in \cW^*_\td(\Lam_0).\end{array}\eqno (2.2.7)$$ 
Here the numerator $\Xi^{\Lam_0,\Om^*_0}\big[\Lam\setminus\Lam_0|\bx (\Lamc )\big]$ is given by 
$$\begin{array}{l}
\diy\Xi^{\Lam_0,\Om^*_0}\big[\Lam\setminus\Lam_0|\bx (\Lamc )\big]\\
\qquad\diy =\int_{\cW_\td(\Lam )}
{\rd}^{\Lam {}}\Om^*_{\Lam\setminus\Lam_0}
{\mathbf 1}\Big(\Om^*_{\Lam\setminus\Lam_0}\in\cW_\td(\Lam\setminus\Lam_0)\Big)\\
\qquad\qquad
\diy\times \alpha_\Lam (\Om^*_{\Lam\setminus\Lam_0})\frac{z^{K(\Om^*_{\Lam\setminus\Lam_0})}}{L(
\Om^*_{\Lam\setminus\Lam_0})}
\exp\;\left[-h\Big(\Om^*_0\vee
\Om^*_{\Lam\setminus\Lam_0}\big|\bx (\Lamc )\Big)\right].\end{array}\eqno (2.2.8)$$
One can say that quantity   $\Xi^{\Lam_0,\Om^*_0}\big[\Lam\setminus\Lam_0|\bx (\Lamc )\big]$
in (2.2.8) represents a partition
function in $\Lam\setminus\Lam_0$  in the external field generated by the CC 
$\bx (\Lamc )$, in presence of a LC $\Om^*_0$ over $\Lam_0$. $\qquad\triangle$ 
\medskip  

{\bf Lemma 2.2} {\sl The PM $\mu_{\Lam|\bx (\Lamc )}$ satisfies the 
following property: $\forall$ $\Lam_0\subset\Lam$ and $\Om^*_0\in \cW^*(\Lam_0)$,
the PDF $f^{\Lam_0}_{\Lam |\bx (\Lamc )}(\Om^*_0)$ introduced in {\rm{(2.2.7)}} has the form
$$f^{\Lam_0}_{\Lam |\bx (\Lamc )}(\Om^*_0)
 \diy ={\mathbf 1}\Big(\Om^*_0\in\cW^*_\td(\Lam_0)\Big)
 \alpha_\Lam (\Om^*_0)\, \frac{z^{K(\Om^*_0)}}{L(\Om^*_0)}
 q^{\Lam_0}_{\Lam |\bx(\Lamc)}(\Om^*_0).\eqno (2.2.9)$$ 
Here $\forall$ $\Lam_0\subseteq\Lam'\subseteq\Lam$, the functional
 $q^{\Lam_0}_{\Lam |\bx(\Lamc)}(\Om^*_0)$ admits the representation  
$$\begin{array}{l}q^{\Lam_0}_{\Lam |\bx(\Lamc)}(\Om^*_0) \\
\diy\qquad =\int_{\cW^*_\td(\Lam )}
 {\rd}\mu_{\Lam|\bx (\Lamc )}(\Om^*_{\Lam\setminus\Lam'})
 {\mathbf 1}\Big(\Om^*_{\Lam\setminus\Lam'}\in \cW^*_\td(\Lam\setminus\Lam' )\Big)\\
\qquad\qquad\qquad\quad\times \alpha_\Lam (\Om^*_{\Lam\setminus\Lam'}) 
\Xi^{\Lam_0,\Om^*_0}\left[\Lam'\setminus\Lam_0 {},\,
\Om^*_{\Lam\setminus\Lam'}\vee\bx (\Lamc )\right].\end{array}\eqno (2.2.10)$$
Moreover, for a given $\Om^*_{\Lam\setminus\Lam'}\in\cW^*_\td(\Lam\setminus\Lam' )$, the conditional partition 
function $\Xi^{\Lam_0,\Om^*_0}\left[\Lam'\setminus\Lam_0 {},\,
\Om^*_{\Lam\setminus\Lam'}\vee\bx (\Lamc )\right]$
is defined in a manner similar to quantity
 $\Xi^{\Lam_0,\Om^*_0}\left[\Lam\setminus\Lam_0 |\bx (\Lamc )\right]$  in {\rm{(2.2.8)}}:
$$\begin{array}{l}
\diy \Xi^{\Lam_0,\Om^*_0}\left[\Lam'\setminus\Lam_0 {},\,
\Om^*_{\Lam\setminus\Lam'}\vee\bx (\Lamc )\right]\\
\diy\quad =\int_{\cW^*_\td(\Lam )}\;
{\rd}^{\Lam {}}\Om^*_{\Lam' \setminus\Lam_0}\;
{\mathbf 1}\Big(\Om^*_{\Lam' \setminus\Lam_0}\in \cW^*_\td(\Lam'\setminus\Lam_0)\Big)
\alpha_\Lam (\Om^*_{\Lam' \setminus\Lam_0})\\  
\diy\qquad\qquad\times
\,\frac{z^{K(\Om^*_{\Lam' \setminus\Lam_0})}}{L(
\Om^*_{\Lam' \setminus\Lam_0})}\,\exp\;\left[-h\Big(\Om^*_0\vee\Om^*_{\Lam' \setminus\Lam_0}\big|
\Om^*_{\Lam\setminus\Lam'}\vee\bx (\Lamc )\Big)\right].\end{array}\eqno (2.2.11)$$ 
 
Furthermore, the RDMK $F^{\Lam_0}_{\Lam |\bx (\Lamc )}$ can be written as an integral: 
$$\begin{array}{l}
\diy F^{\Lam_0}_{\Lam |\bx (\Lamc )}(\bx_0,\by_0)=
\int_{\ucW^*_\td(\bx_0,\by_0)}
{\ubbP}^*_{\bx_0,\by_0}({\rd}\oOm^*_0)\chi^{\Lam_0}(\oOm^*_0)\\
\diy\qquad\qquad\qquad\qquad\qquad\qquad \times\alpha_\Lam (\oOm^*_0)\,z^{K(\oOm^{\,*}_0)}
\;{\wh q}^{\Lam_0}_{\Lam |\bx(\Lamc)}(\oOm^*_0).
\end{array}\eqno (2.2.12)$$
Here the functional ${\wh q}^{\Lam_0}_{\Lam |\bx(\Lamc)}(\oOm^*_0)$ admits the following
representation: $\forall$ $\Lam_0\subseteq\Lam'\subset\Lam$, 
$$\begin{array}{l}
\diy{\wh q}^{\Lam_0}_{\Lam |\bx(\Lamc)}(\oOm^*_0)\\
\diy\qquad =\int_{\cW^*_\td(\Lam )}
{\rd}\mu_{\Lam |\bx (\Lamc )}(\Om^*_{\Lam\setminus\Lam'}){\mathbf 1}\Big(\Om^*_{\Lam\setminus\Lam'}
\in \cW^*_\td(\Lam\setminus\Lam' )\Big)\\
\diy\qquad\times\chi^{\Lam_0}\left(\Om^*_{\Lam\setminus\Lam'}\right)
\alpha_\Lam\left(\Om^*_{\Lam\setminus\Lam'}\right)
{\wh\Xi}^{\Lam_0,\oOm^*_0}\left[\Lam' \setminus\Lam_0\big|\,
\Om^*_{\Lam\setminus\Lam'}\vee\bx (\Lamc )\right]\,.\end{array}\eqno (2.2.13)$$
Moreover, in analogy with {\rm{(2.2.11)}},  for a given} $\Om^*_{\Lam\setminus\Lam'}\in\cW^*_\td
(\Lam\setminus\Lam' )$,
$$\begin{array}{l}
\diy{\wh\Xi}^{\Lam_0,\oOm^*_0}\left[\Lam' \setminus\Lam_0\big|\,
\Om^*_{\Lam\setminus\Lam'}\vee\bx (\Lamc )\right]\\
\diy \;\;\; =\int_{\cW^*_\td(\Lam )}
{\rd}^\Lam\Om^*_{\Lam' \setminus\Lam_0}\;
{\mathbf 1}\Big(\Om^*_{\Lam' \setminus\Lam_0}\in \cW^*_\td(\Lam'\setminus\Lam_0)\Big)
\chi^{\Lam_0}(\Om^*_{\Lam' \setminus\Lam_0})\\
\diy\;\;\times \alpha_\Lam (\Om^*_{\Lam' \setminus\Lam_0})
\,\frac{z^{K(\Om^*_{\Lam' \setminus\Lam_0})}}{L(
\Om^*_{\Lam' \setminus\Lam_0})}\,\exp\;\left[-h\Big(\oOm^*_0\vee
\Om^*_{\Lam' \setminus\Lam_0}\big|
\Om^*_{\Lam\setminus\Lam'}\vee\bx (\Lamc )\Big)\right].\end{array}\eqno (2.2.14)$$
$\Box$

\medskip

\medskip

Note that the presence of terms $\exp\;\left[-h\Big(\Om^*_0\vee
\Om^*_{\Lam' \setminus\Lam_0}\big|
\Om^*_{\Lam\setminus\Lam'}\vee\bx (\Lamc )\Big)\right]$ and 
$\exp\;\left[-h\Big(\uOm^*_0\vee
\Om^*_{\Lam' \setminus\Lam_0}\big|
\Om^*_{\Lam\setminus\Lam'}\vee\bx (\Lamc )\Big)\right]$ in (2.2.11) and (2.2.14) 
implies the presence of the indicators ${\mathbf 1}\Big(\Om^*_0\vee\Om^*_{\Lam'\setminus\Lam_0}
\vee\Om^*_{\Lam\setminus\Lam'}\in\cW^*_\td(\Lam)\Big)$ and\\ 
${\mathbf 1}\Big((\uOm^*_0,\Om^*_{\Lam'\setminus\Lam_0}
\vee\Om^*_{\Lam\setminus\Lam'})\in\ucW^{\,*}_\td(\Lam_0,\Lam\setminus\Lam_0)\Big)$.

In particular, for $\Lam_0=\Lam'$, Eqns (2.2.10) and  (2.2.13) take the form:
$$\begin{array}{l}q^{\Lam_0}_{\Lam |\bx(\Lamc)}(\Om^*_0) \\
\diy\qquad =\int_{\cW^*_\td(\Lam )}
 {\rd}\mu_{\Lam|\bx (\Lamc )}(\Om^*_{\Lam\setminus\Lam_0})
 {\mathbf 1}(\Om^*_{\Lam\setminus\Lam_0}\in \cW^*_\td(\Lam\setminus\Lam_0))\qquad{}\\
\qquad\qquad\qquad\qquad\times \alpha_\Lam (\Om^*_{\Lam\setminus\Lam_0}) 
\exp\;\left[-h\Big(\Om^*_0\big|
\Om^*_{\Lam\setminus\Lam_0}\vee\bx (\Lamc )\Big)\right]\end{array}\eqno (2.2.15)$$
and
$$\begin{array}{l}
\diy{\wh q}^{\Lam_0}_{\Lam |\bx(\Lamc)}(\uOm^*_0)\\
\diy\qquad =\int_{\cW^*_\td(\Lam )}
{\rd}\mu_{\Lam |\bx (\Lamc )}(\Om^*_{\Lam\setminus\Lam_0}){\mathbf 1}(\Om^*_{\Lam\setminus\Lam_0}
\in \cW^*_\td(\Lam\setminus\Lam_0))\\
\diy\qquad\times\chi^{\Lam_0}\left(\Om^*_{\Lam\setminus\Lam_0}\right)
\alpha_\Lam\left(\Om^*_{\Lam\setminus\Lam_0}\right)
\exp\;\left[-h\Big(\uOm^*_0\big|
\Om^*_{\Lam\setminus\Lam_0}\vee\bx (\Lamc )\Big)\right]
\,.\end{array}\eqno (2.2.16)$$

On the other hand, when $\Lam'=\Lam$, Eqn (2.2.13) coincides with (2.2.4).
\medskip

We would like to stress here that the integral $\diy\int
 {\rd}\mu_{\Lam|\bx (\Lamc )}(\Om^*_{\Lam\setminus\Lam'})$ in (2.2.9), (2.2.13) is taken 
 in the variable $\Om^*_{\Lam\setminus\Lam'}$ considered as an element of space
 $\cW^*_\td(\Lam )$.   Likewise, the integral $\diy\int
{\rd}^\Lam\Om^*_{\Lam' \setminus\Lam_0}$ in (2.2.11), (2.2.14)
is taken 
 in the variable $\Om^*_{\Lam' \setminus\Lam_0}$ considered as an element of space
 $\cW^*_\td(\Lam )$. 
\medskip
 
Eqns  (2.2.9)--(2.2.14) are called the FK-DLR equations in volume $\Lam$. 

\medskip

\medskip

{\bf 2.3. The infinite-volume FK-DLR equations and RDMKs.}\\
Infinite-volume versions of the RDMK arise when we mimic properties listed
in Lemmas 2.1 and 2.2 by getting rid of the reference to the enveloping cube $\Lam$ (this includes
the external CC $\bx (\Lamc )$ and the functional $\alpha_\Lam$ 
indicating Dirichlet's boundary condition).
The first place to do so is the PM $\mu_{\Lam|\bx (\Lamc )}$; 
to this end we need to consider its infinite-volume analog  $\mu_{\bbR^d}$ representing
an RMPP in the whole plane $\bbR^d$. Formally, $\mu_{\bbR^d}$ yields a PM
on the sigma-algebra $\fW\left(\bbR^d\right)$ of subsets in $\cW^*\left(\bbR^d\right)$.
The space   $\cW^*\left(\bbR^d\right)$ is formed by pairs $\left[\bx (\bbR^d),\bOm^*(\bx(\bbR^d))\right]$ 
where $\bx (\bbR^d)$ is a locally finite set in the plane and $\bOm^*(\bx(\bbR^d))$
(in brief, $\bOm^*_{\bbR^d}$ or simply $\bOm^*$) is a collection $\{\om^*(x),\;x\in\bx (\bbR^d)\}$
of loops $\om^*(x)\in \cW^*_x$. Alternatively, $\bOm^*(\bx (\bbR^d))\in
{\operatornamewithlimits{\times}\limits_{x\in\bx (\Lam )}}\cW^*_x$. Next, $\fW\left(\bbR^d\right)$ 
is the sigma-algebra 
of subsets in  $\cW^*\left(\bbR^d\right)$ generated by the cylinder events. It is convenient 
to refer to $\fW\left(\bbR^d\right)$ as the smallest sigma-algebra containing all sigma-algebras
$\fW (\Lam )$ where $\Lam$ is a cube and $\fW (\Lam )$ is formed by 
the inverse images of sets $A\in\fW_\Lam$ under the maps  
$$\bx (\bbR^d)\mapsto\bx (\Lam ),\;\;  \bOm^*_{\bbR^d}\mapsto
\bOm^*_\Lam .$$ 
Here $\bx (\Lam )=\bx (\bbR^d)\cap\Lam$ and, for  $\bOm^*_{\bbR^d}
=\{\om^*(x),\;x\in\bx (\bbR^d)\}$, the symbol
$\bOm^*_{\Lam }$ stands for the sub-collection $\{\om^*(x),\;x\in\bx (\Lam )\}$.  

To simplify technical aspects of the presentation, we will omit the reference to the initial
CC $\bx (\bbR^d)$ and write $\bOm_{\bbR^d}\in\cW^*(\bbR^d)$ or $\bOm^*\in\cW^*(\bbR^d)$
(given a LC  $\bOm^*$, the initial CC is uniquely determined and 
can be denoted by $\bx (\bOm^*)$). 

Furthermore, we will use the notation $\cW^*(\Lamc)$ for the subset in
$\cW^*(\bbR^d)$ formed by LCs 
 $\bOm^*_{\Lamc}$ with 
 $\bx (\bOm^*_{\Lamc})\in\cC_\td(\Lamc)$. (We call 
 such $\bOm^*_{\Lamc}$ a LC over $\Lamc$.)

\medskip

\medskip

{\bf Definition 2.5.} We say that a PM $\mu =\mu_{\bbR^d}$ on
$\left(\cW^*\left(\bbR^d\right),\fW\left(\bbR^d\right)\right)$ satisfies the (infinite-volume)
FK-DLR equations if the restriction $\mu\Big|_{\fW^*(\Lam_0)}$ of $\mu$ to  $\fW^*(\Lam_0)$
is given by the PDF $ f^{\Lam_0}(\Om^*_0):=\diy\frac{\mu\Big|_{\fW^*(\Lam_0)}({\rd}\Om^*_0)}{{\rd}^{\Lam_0} \Om^*_0}\;$, $\Om^*_0\in \cW^*(\Lam_0)$, of the form  
$$\diy f^{\Lam_0}(\Om^*_0)
={\mathbf 1}\Big(\Om^*_0\in\cW^*_\td(\Lam_0)\Big)\frac{z^{K(\Om^*_0)}}{L(\Om^*_0)}\,q^{\Lam_0}(\Om^*_0),
\eqno (2.3.1)$$
where the functional $q^{\Lam_0}(\Om^*_0)$ admits the following representation. $\forall$ pair of cubes
$\Lam_0\subset\Lam\subset\bbR^d$,
$$\begin{array}{l}
q^{\Lam_0}(\Om^*_0) =\diy\int_{\cW^*_\td(\bbR^d)}{\rd}\mu (\bOm^*_{\Lamc})
{\mathbf 1}\Big(\bOm^*_{\Lamc}\in\cW^*_\td(\Lamc )\Big)\\
\qquad\qquad\qquad\qquad\qquad\qquad\times\Xi^{\Lam_0,\Om^*_0}\left(\Lam\setminus\Lam_0\big|\bOm^*_{\Lamc}\right)\,.\end{array}\eqno (2.3.2)$$
Observe similarities with Eqn (2.2.9). At the same time,  note the absence of 
the indicator $\alpha_\Lam$ 
in the RHS of (2.3.2). Here, for a given (infinite) LC 
$\bOm^*_{\Lamc}\in \cW^*_\td(\Lamc )$, 
the expression $\Xi^{\Lam_0,\Om^*_0}\left(\Lam\setminus\Lam_0\big|\bOm^*_{\Lamc}\right)$
yields a partition function in $\Lam \setminus\Lam_0$, in the external 
field generated by  $\bOm^*_{\Lamc}$ and in presence of
a LC $\Om^*_0\in \cW^*_\td(\Lam_0)$:
$$\begin{array}{l}\diy\Xi^{\Lam_0,\Om^*_0}\left(\Lam\setminus\Lam_0\big|\bOm^*_{\Lamc}
\right) =\diy
\int_{\cW^*_\td(\Lam\setminus\Lam_0)}{\rd}\Om^*_{\Lam\setminus\Lam_0}
\frac{z^{K(\Om^*_{\Lam \setminus\Lam_0})}}{L(
\Om^*_{\Lam \setminus\Lam_0})}\\
\diy\qquad\times\,{\mathbf 1}\Big(\Om^*_0\vee\Om^*_{\Lam\setminus\Lam_0}\vee
\bOm^*_{\Lamc}\in\cW^*_\td(\bbR^d)\Big)\\
\qquad\qquad\qquad\qquad\qquad\qquad\times\exp\;\left[-h\left(\Om^*_0\vee\Om^*_{\Lam\setminus\Lam_0}\big|
\bOm^*_{\Lamc}\right) \right].\end{array}\eqno (2.3.3)$$
Comparing to Eqn (2.2.10) we see a difference: the integral $\diy\int_{\cW^*_\td(\Lam\setminus\Lam_0)}{\rd}\Om^*_{\Lam\setminus\Lam_0}$ in (2.3.3) provides a simplification. In turn, $h\left(\Om^*_0\vee\Om^*_{\Lam\setminus\Lam_0}\big|
\bOm^*_{\Lamc}\right)$ represents the energy of the concatenated LC 
$\Om^*_0\vee\Om^*_{\Lam \setminus\Lam_0}$ over $\Lam$,
in the external potential generated by the LC 
$\bOm^*_{\Lamc}$ over $\Lamc$. Formally,  
$h\left(\Om^*_0\vee\Om^*_{\Lam\setminus\Lam_0}\big|
\bOm^*_{\Lamc}\right)$ is defined, for $\Om^*_0\vee\Om^*_{\Lam\setminus\Lam_0}
\vee\bOm^*_{\Lamc}\in\cW^*_\td(\bbR^d)$, as the limit:
$$\begin{array}{l}h\left(\Om^*_0\vee\Om^*_{\Lam
\setminus\Lam_0}\big|\bOm^*_{\Lamc}\right) =\lim\limits_{L\to +\infty}
h\left[\Om^*_0\vee\Om^*_{\Lam \setminus\Lam_0}\big|
\bOm^*_{\Lam (L)\setminus\Lam}\right].\end{array}\eqno (2.3.4)$$
Here $\Lam (L)$ stands for the cube $[-L,L]^{\times d}$ of side-length $2L$ 
centered at the origin in $\bbR^d$ and $\bOm^*_{\Lam (L)\setminus\Lam}$ denotes
the restriction $\bOm^*_{\Lamc}\Big|_{\Lam (L)\setminus\Lam}$ of 
$\bOm^*_{\Lamc}$ to $\Lam (L)\setminus\Lam$. The equations (2.3.1)--(2.3.4) 
are referred to as infinite-volume FK-DLR equation.

For short, a PM $\mu =\mu_{\bbR^d}$ satisfying (2.3.1)--(2.3.4) is called FK-DLR; we will also 
employ the term an FK-DLR probability measure (FK-DLR PM). The class of 
FK-DLR PMs (for a given pair of values $z\in (0,1),\beta\in (0,+\infty )$) 
is denoted by $\fK (z,\beta )$, or, briefly, $\fK$.  It is straightforward that
any PM $\mu\in\fK$ is supported by the set $\cW^*_\td(\bbR^d)$: $\mu (\cW^*_\td(\bbR^d))=1$. 
$\qquad\triangle$

\medskip   

\medskip

{\bf Definition 2.6.} Let $\mu\in\fK (z,\beta )$ be an FK-DLR PM. In this definition
we associate with $\mu$ 
a family of integral kernels $F^{\Lam_0}(\bx_0,\by_0)$ where $\bx_0,\by_0\in
{\cC}(\Lam_0)$ and $\Lam_0\subset\bbR^d$ is an 
arbitrary cube. 
Namely, when $\sharp\;\bx_0=\sharp\;\by_0$, we set:
$$\diy F^{\Lam_0}(\bx_0,\by_0)=\int_{\ucW^*_\td(\bx_0,\by_0)}
{\rd}{\ubbP}^*_{\bx_0,\by_0}(\uOm^*_0)\;z^{K(\uOm^*_0)}\;
\chi^{\Lam_0}(\uOm^*_0)\;{\wh q}^{\,\Lam_0}(\uOm^*_0).\eqno (2.3.5)$$
In turn, the quantity ${\wh q}^{\,\Lam_0}(\uOm^*_0)$ admits the following integral representation
involving PM $\mu$: 
$\forall$ cube $\Lam\subset\bbR^d$ containing $\Lam_0$,
$$\begin{array}{l}
\diy {\wh q}^{\,\Lam_0}(\uOm^*_0)=\int_{\cW^*_\td(\bbR^d)}
{\rd}\mu (\bOm^*_{\Lamc})\chi^{\Lam_0}(\bOm^*_{\Lamc})\\ 
\qquad\times{\mathbf 1}\Big((\uOm^{\,*}_0,\bOm^*_{\Lamc})
\in\ucW^{\,*}_\td(\Lam_0,\Lamc)\Big)\,
{\wh\Xi}^{\Lam_0,\uOm^*_0}\left(\Lam\setminus\Lam_0\big|\bOm^*_{\Lamc}\right).
\end{array}\eqno (2.3.6)$$
Like before, $\ucW^{\,*}_\td(\Lam_0,\Lamc)$ denotes here the subset of the Cartesian product\\
$\ucW^{\,*}_\td(\Lam_0)\times\cW^*_\td(\Lamc)$ formed by pairs $(\uOm^{\,*}_0,\bOm^*_{\Lamc})$
such that $\forall$ path $\oom^*\in\uOm^{\,*}_0$ and $0\leq {\ov l}<k(\oom^*)$ 
and $\forall$ $\om^*\in\bOm^*_{\Lamc}$ and $0\leq l<k(\om^*)$, the Euclidean norm 
$$|\oom^*({\ov l}\beta +\ttt )-\om^*(l\beta +\ttt  )|_{\rm{Eu}}>\td\;\;\forall\;\;t\in [0,\beta ]. $$
Next, for a given PC $\uOm^{\,*}_0\in\ocW^{\,*}_\td(\Lam_0)$ over $\Lam_0$ and  LC $\bOm^*_{\Lamc}\in 
\cW^*_\td(\Lamc )$ over $\Lamc$ such that the pair $(\uOm^{\,*}_0,\bOm^*_{\Lamc})
\in\ocW^{\,*}_\td(\Lam_0,\Lamc)$, the expression 
${\wh\Xi}^{\Lam_0,\uOm^*_0}
\left(\Lam\setminus\Lam_0\big|\bOm^*_{\Lamc}\right)$ is defined similarly to (2.2.11):
$$\begin{array}{r}\diy{\wh\Xi}^{\Lam_0,\uOm^*_0}
\left(\Lam\setminus\Lam_0\big|\bOm^*_{\Lamc}\right)
=\int_{\cW^*_\td(\Lam \setminus\Lam_0)}{\rd}\Om^*_{\Lam\setminus\Lam_0}
\chi^{\Lam_0}(\Om^*_{\Lam\setminus\Lam_0})\qquad\qquad{}\\
\times{\mathbf 1}\Big( (\uOm^*_0,\Om^*_{\Lam\setminus\Lam_0}\vee\bOm^*_{\Lamc})\in
\ocW^{\,*}_\td(\Lam_0,\bbR^d\setminus \Lam)\Big)\qquad{}\\
\diy\times\frac{z^{K(\Om^*_{\Lam\setminus\Lam_0})}}{L(
\Om^*_{\Lam\setminus\Lam_0})}\exp\;\left[-h\left(\uOm^*_0\vee\Om^*_{\Lam\setminus\Lam_0}\big|
\bOm^*_{\Lamc}\right) \right].\end{array}\eqno (2.3.7)$$
Again, $\ocW^{\,*}_\td(\Lam_0,\bbR^d\setminus \Lam)$ stands for the subset in 
$\ocW^{\,*}_\td(\Lam_0)\times\cW^*_\td(\bbR^d\setminus \Lam)$ formed by pairs
$(\uOm^*_0,\bOm^*_{\bbR^d\setminus \Lam})$ such that 
$\forall$ path $\oom^*\in\uOm^{\,*}_0$ and $0\leq {\ov l}<k(\oom^*)$ 
and $\forall$ $\om^*\in\bOm^*_{\bbR^d\setminus \Lam}$ and $0\leq l<k(\om^*)$, 
$$|\oom^*({\ov l}\beta +\ttt )-\om^*(l\beta +\ttt  )|_{\rm{Eu}}>a\;\;\forall\;\;t\in [0,\beta ]. $$
The indicators $\chi^{\Lam_0}(\uOm^*_0)$,  $\chi^{\Lam_0}(\obOm^*_{\Lamc})$ and 
$\chi^{\Lam_0}(\Om^*_{\Lam\setminus\Lam_0})$  in (2.3.5)--(2.3.7) are defined similarly to (2.2.5)
(and play a similar role). Namely, for a PC 
$\;\uOm^*_0=\{\oom^*\}\in\ocW^{\,*}(\bx_0,\by_0)$ and an LC
$\;\Om^*_{\Lam\setminus\Lam_0}=\{\om^*_x,x\in\bx_{\Lam\setminus\Lam_0}\}\in \cW^*(\Lam\setminus\Lam_0)$,
with $\bx_{\Lam\setminus\Lam_0}\in\cC (\Lam\setminus\Lam_0)$:
$$\chi^{\Lam_0}(\uOm^*_0)={\mathbf 1}\Big(\oom^*({\ov l}\beta )\in\bbR^d
\setminus\Lam_0,\;\forall\;1\leq {\ov l}<k\left(\oom^*\right)\;\forall\;\oom^*\in
\uOm^*_0 \Big),$$
and
$$\chi^{\Lam_0}(\Om^*_{\Lam\setminus\Lam_0})={\mathbf 1}\Big(\om^*_x(l\beta )
\in\bbR^d\setminus \Lam\;\forall\;l< k\left(\om^*_x\right)\;\forall\;x\in
\bx(\Om^*_{\Lam\setminus\Lam_0}) \Big).$$
Likewise, for a LC $\bOm^*_{\Lamc}=\{\om^*_x,x\in\bx (\bOm^*_{\Lamc})\}\in\cW^*(\Lamc )$:
$$\begin{array}{l}
\chi^{\Lam_0}(\bOm^*_{\Lamc})={\mathbf 1}\Big(\om^*_x (l\beta )\in\bbR^d\setminus \Lam\;\forall\;l< k(\om^*_x),\om^*_x\in
\bOm^*_{\Lamc}\Big).\end{array}\eqno (2.3.8)$$
Finally, similarly to (2.3.4), for $(\uOm^*_0,\Om^*_{\Lam\setminus\Lam_0}\vee\bOm^*_{\Lamc})\in
\ocW^{\,*}_\td(\Lam_0,\bbR^d\setminus \Lam)$  we set:
$$\begin{array}{l}h\left(\uOm^*_0\vee\Om^*_{\Lam
\setminus\Lam_0}\big|
\bOm^*_{\Lamc}\right)
=\lim\limits_{L\to +\infty}
h\left[\uOm^*_0\vee\Om^*_{\Lam \setminus\Lam_0}\big|
\bOm^*_{\Lam (L)}\right].\end{array}\eqno (2.3.9)$$
\medskip

When $\sharp\;\bx_0\neq\sharp\;\by_0$, we set: $F^{\Lam_0}(\bx_0,\by_0)=0$. $\qquad\triangle$
\medskip

It is instructive to re-write the definitions (2.3.4) and (2.3.9) in line with (2.1.2), (2.1.4), 
(2.1.8)  and (2.1.10), 
expressing the functionals\\ $h\left(\Om^*_0\vee\bOm^*_{\Lam
\setminus\Lam_0}\big|\bOm^*_{\Lamc}\right)$ and 
$h\left(\uOm^*_0\vee\Om^*_{\Lam
\setminus\Lam_0}\big|\bOm^*_{\Lamc}\right)$ in terms of energies of 
CCs $\Om^*_0(\ttt )\vee\Om^*_{\Lam
\setminus\Lam_0}(\ttt )$, $\uOm^*_0(\ttt )\vee\Om^*_{\Lam\setminus\Lam_0}(\ttt )$
and $\bOm^*_{\Lamc}(\ttt )$ forming $\ttt$-sections of the 
corresponding PCs and LCs, where $0\leq \ttt\leq\beta$. Namely,
$$h\left(\Om^*_0\vee\Om^*_{\Lam\setminus\Lam_0}|\bOm^*_{\Lamc}\right)
=\int_0^\beta \rd\ttt E\left[\Om^*_0(\ttt )\vee\Om^*_{\Lam\setminus\Lam_0}(\ttt )|
\bOm^*_{\Lamc}(\ttt )\right]\eqno (2.3.10)$$
where 
$$\begin{array}{l}E\left[\Om^*_0(\ttt )\vee\Om^*_{\Lam\setminus\Lam_0}(\ttt )|\bOm^*_{\Lamc}(\ttt )\right]\\
\qquad =E\left[\Om^*_0(\ttt )\vee \Om^*_{\Lam
\setminus\Lam_0}(\ttt )\right]+E\left[ \uOm^*_0(\ttt )\vee\Om^*_{\Lam
\setminus\Lam_0}(\ttt )||\bOm^*_{\Lamc}(\ttt )\right]
\end{array}\eqno (2.3.11)$$
and
$$\begin{array}{l}E\left[\uOm^*_0(\ttt )\vee\Om^*_{\Lam\setminus\Lam_0}(\ttt )|\bOm^*_{\Lamc}(\ttt )\right]\\
\qquad =E\left[\uOm^*_0(\ttt )\vee \Om^*_{\Lam\setminus\Lam_0}(\ttt )\right]+E\left[ \uOm^*_0(\ttt )\vee\Om^*_{\Lam
\setminus\Lam_0}(\ttt )||\bOm^*_{\Lamc}(\ttt )\right].
\end{array}\eqno (2.3.12)$$
In turn, finite CCs $\Om^*_0(\ttt )$, $\uOm^*_0(\ttt )$ and $\Om^*_{\Lam\setminus\Lam_0}(\ttt )$
and an infinite CC $\bOm^*_{\Lamc}(\ttt )$ are given by
$$\Om^*_0(\ttt )=\operatornamewithlimits{\cup}\limits_{\om^*\in\Om^*_0}
\operatornamewithlimits{\cup}\limits_{0\leq l<k(\om^*)}\{\om^*(l\beta +\ttt )\},\;
\uOm^*_0(\ttt )=\operatornamewithlimits{\cup}\limits_{\oom^*\in\uOm^*_0}
\operatornamewithlimits{\cup}\limits_{0\leq l<k(\oom^*)}\{\oom^*(l\beta +\ttt )\},$$
$$\Om^*_{\Lam\setminus\Lam_0}(\ttt )=\operatornamewithlimits{\cup}\limits_{\om^*\in\Om^*_{\Lam\setminus\Lam_0}}
\operatornamewithlimits{\cup}\limits_{0\leq l<k(\om^*)}\{\om^*(l\beta +\ttt )\}$$
and
$$\begin{array}{cl}\bOm^*_{\Lamc}(\ttt )&=\operatornamewithlimits{\bigcup}\limits_{\diy\om^*\in\bOm^*_{\Lamc}}
\operatornamewithlimits{\cup}\limits_{0\leq l<k(\oom^*)}\{\om^*(l\beta +\ttt )\}\\
\;&=\Big\{\om^*(l\beta +\ttt ):\;\om^*\in\bOm^*_{\Lamc},\;0\leq l<k(\oom^*)\Big\}.
\end{array}$$

Owing to the FK-DLR property of $\mu$, the RHS in (2.3.6) does not depend
on the choice of the cube $\Lam\supset\Lam_0$. Moreover, the kernels
$F^{\Lam_0}$ satisfy the compatibility property: $\forall$ pair
of cubes $\Lam_1\subset\Lam_0$,
$$\int_{{\cC}(\Lam_0\setminus\Lam_1)}{\rd}\bz\, F^{\Lam_0}(\bx_1\vee\bz,
\by_1\vee\bz)=F^{\Lam_1}(\bx_1,\by_1),\;\;\bx_1,\by_1\in{\cC}(\Lam_0).
\eqno (2.3.13)$$
In particular, 
$$\int_{{\cC}(\Lam_0)}{\rd}\bz\,F^{\Lam_0}(\bz, \bz )=1.\qquad\triangle\eqno (2.3.14)$$
 
\medskip

{\bf Definition 2.7.} Let $\mu$ be FK-DLR and $\{F^{\Lam_0}\}$ be the family 
of  kernels  associated with $\mu$ by Eqns   (2.3.4)--(2.3.7). Given 
$\phi_{\Lam_0}\in\cH (\Lam_0)$, introduce a trace-class operator $R^{\Lam_0}$ acting in $\cH(\Lam_0)$:
$$R^{\Lam_0}\phi_{\Lam_0}(\bx_0)=\int_{{\cC}(\Lam_0)}
{\rd}\by_0\,F^{\Lam_0}(\bx_0,\by_0)\phi_{\Lam_0}(\by_0),\;\;\bx_0\in{\cC}
(\Lam_0).\eqno (2.3.15)$$
Then, according to (2.3.13)--(2.3.14), 
$${\tr}_{\cH (\Lam_0\setminus\Lam_1)}R^{\Lam_0}=R^{\Lam_1},\;\;
{\tr}_{\cH (\Lam_0)}R^{\Lam_0}=1.\eqno (2.3.16)$$
The family of operators $R^{\Lam_0}$ defines a linear normalized functional 
on the quasilocal C$^*$-algebra $\fB (\bbR^d)$ such that for $A\in\fB(\Lam_0)$
$$\varphi (A)=\tr_{\cH (\Lam_0)}\big(AR^{\Lam_0}\big).\eqno (2.3.17)$$
We call it the FK-DLR functional generated by $\mu$; to stress this fact
we sometimes use the notation $\varphi_\mu$. If in addition
$\varphi$ is a state (that is, the operators $R^{\Lam_0}$ are positive-definite)
then we say that $\varphi$ is an FK-DLR state. In this case we 
call the operator $R^{\Lam_0}$ an infinite-volume FK-DLR RDM. The
class of FK-DLR functionals is denoted by $\fF=\fF (z,\beta )$ and its subset consisting 
of the FK-DLR states by $\fF_+=\fF_+(z,\beta )$. $\qquad\triangle$
\medskip

Before we move further, we would like to introduce a property conventionally called
a Ruelle bound. This is closely related to the so-called Campbell theorem
assessing integrals of summatory functions $\Sigma_g:\bOm^*\in\cW^*_\td(\bbR^d)\mapsto
\sum\limits_{\Om^*\subset\bOm^*}g(\Om^*)$:
$$\int_{\cW^*_\td(\bbR^d)}\mu (\rd\bOm^*)\Sigma_g(\bOm^*)=
\int\rd\Om^*\rho (\Om^*)g(\Om^*)\eqno (2.3.18)$$
where $\rho =\rho_\mu$ is the moment function of the RMPP $\mu$. The Ruelle bound
with a constant $\orho =z\exp (\beta{\ov V}\tR^d/\td^d)$ (cf. (1.1.19)) reads
$$\rho (\Om^*)\leq\frac{\orho^{K(\Om^*)}}{L(\Om^*)}\eqno (2.3.19)$$
and follows from the representation
$$\rho (\Om^*)=\frac{z^{K(\Om^*)}}{L(\Om^*)}
\int_{\cW^*_\td(\bbR^d)}\mu (\rd\bOm^*)\exp\,\big[-h(\Om^*|\bOm^*) \big]\eqno (2.3.20)$$
and the lower estimate 
$$h(\Om^*|\bOm^*)\geq -\beta{\ov V}\frac{\tR^d}{\td^d}K(\Om^*).\eqno (2.3.21)$$
In turn, the bound (2.3.21) is deduced from representations (2.3.10)--(2.3.12), the definition of $\ov V$
(see Eqn (1.1.4)) and the observation that, under the hard-core and the finite-range assumptions
on the two-body potential $V$, the (classical) energy of interaction between a particle   
and a CC is always $\geq -{\ov V}\tR^d/\tr^d$.

The bound (2.3.19) will play a r\^{o}le in arguments conducted in Section 3 and in paper 
\cite{SKS}.
\medskip

\medskip

{\bf 2.4. Results on infinite-volume FK-DLR PMs, functionals and states.} Our results
about the classes $\fK$, $\fF$ and $\fF_+$ are summarized in the following theorems.
\medskip

{\bf Theorem 2.1.} {\sl The class $\fK (z,\beta )$ of FK-DLR
PMs is non-empty. Moreover, the family of FK-DLR PMs $\mu_\Lam$ is compact
in the weak topology, and every limiting point $\mu$ for this family lies in $\fK (z,\beta )$. Furthermore,
the family of the Gibbs states $\varphi_\Lam$ is compact in the w$^*$-topology, and every 
limiting point for this family gives an element from $\fF_+$.  The same is true for any family 
of the PMs $\mu_{\Lam |\bx (\Lamc)}$ and states $\varphi_{\Lam |\bx (\Lamc )}$ with 
external CCs $\bx (\Lamc)\in{\cC}_\td(\Lamc)$. Consequently, the set $\fF_+(z,\beta )$ is non-empty.}
$\qquad\Box$

\medskip

As above, in dimension two we state a result on shift-invariance to be proved in \cite{SKS}.
We adopt the conditions and notations from Theorem 1.2. 

\medskip

\medskip

{\bf Remark.} As was noted earlier, the formalism developed in this paper does not cover
the phenomenon of Bose--Einstein condensation. A frequently expressed opinion is that
it requires ensembles loops and trajectories of an infinite time-length.  
\medskip

\medskip

{\bf Theorem 2.2.} {\sl Take $d=2$ and let $\mu$ be a PM
from $\fK (z,\beta )$. Then the corresponding FK-DLR functional $\varphi_{\mu}\in\fF(z,\beta )$
is shift-invariant: $\forall$ square $\Lam_0\subset\bbR^{2}$, vector $s\in\bbR^{2}$ and 
operator $A\in\fB (\Lam_0)$,
$$\varphi_\mu (A)=\varphi_\mu ({\tS}(s)A),\;\;A\in\fB(\bbR^{2}).$$
In terms of the corresponding infinite-volume RDMs} $R^{\Lam_0}$:
$$R^{\tS(s)\Lam_0} ={\tU}^{\Lam_0}(s)R^{\Lam_0}\,{\tU}^{\Lam_0}(-s).$$
$\Box$ 

\vskip 1 truecm

\section{Proof of Theorems 1.1 and 2.1: a\\ compactness argument}

\medskip

\medskip

Let us fix a cube $\Lam_0$ of side length $2L_0$ centered at $c=(\tc^1,\ldots ,\tc^d)$: 
cf. Eqn (1.1.16).  The first step in the proof is to verify that, as $\Lam_0\subset\Lam$ and cube 
$\Lam\nearrow\bbR^d$,  the RDMK $F^{\Lam_0}_{\Lam |\bx (\Lamc )}(\bx_0,\by_0)$ 
(see (2.2.2)--(2.2.4), (2.2.12)--(2.2.16)) form a compact family in 
$C^0(\cC_\td(\Lam_0)\times\cC_\td(\Lam_0))$. 
(Recall, we work with pairs $(\bx_0,\by_0)$ with $\sharp\;\bx_0=\sharp\;\by_0$.) Note that 
Cartesian product
$\cC_\td(\Lam_0)\times\cC_\td(\Lam_0)$, the range of variable $(\bx_0,\by_0)$,  is compact. 
In fact, set: $v_0=\Big\lceil(2L_0)^d\big/\td^d\Big\rceil$; then any CC $\bx_0\in\cC_\td(\Lam_0)$
 must have $\sharp\;\bx_0\leq v_0$. As in \cite{KS1}, \cite{KS2} and \cite{KSY2},
it is convenient to employ the Ascoli--Arzela theorem, i.e., verify that the
functions $F^{\Lam_0}_{\Lam |\bx (\Lamc )}(\bx_0,\by_0)$ are uniformly bounded and 
equi-continuous. 

Checking uniform boundedness is straightforward: from (2.2.12)--(2.2.16) one can see that
the functional ${\wh q}^{\Lam_0}_{\Lam |\bx(\Lamc)}$  satisfies
$$\begin{array}{l}
\diy{\wh q}^{\Lam_0}_{\Lam |\bx(\Lamc)}(\uOm^*_0)
\leq\exp\,\Big[\beta K(\uOm^*_0){\ov V}\,\tR^d/\td^d\Big]
\end{array}\eqno (3.1)$$ 
cf. (1.1.4). By using (2.2.12)-(2.2.16), we obtain that  $\forall$ $(\bx_0,\by_0)\in \cC_\td(\Lam_0)\times\cC_\td(\Lam_0)$, 
$$F^{\Lam_0}_{\Lam |\bx (\Lamc )}(\bx_0,\by_0)
\leq v_0!\left(1\vee\sum\limits_{k\geq 1}
\orho^k\Big/(2\pi\beta k)^{d/2}\right)^{v_0}\eqno (3.2)$$ 
which yields uniform boundedness in view of (1.1.19).\medskip\medskip

The argument for  equi-continuity of RDMKs is based on uniform bounds upon the gradients
$\nabla_xF^{\Lam_0}_{\Lam |\bx (\Lamc )}(\bx_0,\by_0)$ and
$\nabla_yF^{\Lam_0}_{\Lam |\bx (\Lamc )}(\bx_0,\by_0)$, for $x\in\bx_0$, $y\in\by_0$.
Both cases are treated in a similar fashion; for definiteness, we consider
gradients  $\nabla_yF^{\Lam_0}_{\Lam |\bx (\Lamc )}(\bx_0,\by_0)$.

It can be seen from representation (2.2.12)--(2.2.16) that there are two 
contributions into the gradient. The first contribution comes from varying the measure
${\ubbP}^*_{\;\bx_0,\by_0}$.  The second
one emerges from varying the functional 
${\wh q}^{\Lam_0}_{\Lam |\bx (\Lamc )}(\uOm^*_{\,0})$, more precisely,
the numerator 
${\wh\Xi}^{\Lam_0,\uOm^*_{\,0}}\big[\Lam\setminus\Lam_0 |\bx (\Lamc )\big]$
in (2.2.3). In fact, it is clear that the second contribution will come out only when we vary 
the term $\exp\;\left[
-h\Big(\uOm^*_{\,0}\vee\Om^*_{\Lam\setminus\Lam_0}\big|\bx (\Lamc )\Big)\right]$ in (2.2.4).
Of course, we are interested in variations of a chosen point $y\in\by_0$.

Suppose the CCs are $\bx_0=(x(1),\ldots ,x(n))$ and $\by_0=(y(1),\ldots ,y(n))$ and 
the PC is $\oOm^*_0=(\oom^*(1),\ldots ,\oom^*(n))$.
Effectively, we have to analyze the gradient $\nabla_{y(j)}$ of the following expression:
$$\begin{array}{l}
\diy \int_{\ocW^*_{\,\td}(\bx_0,\by_0)}
{\obbP}^*_{\;\bx_0,\by_0}({\rd}\oOm^*_0)\exp\;\left[-h\Big(\oOm^*_0\big|
\Om^*_{\Lam\setminus\Lam_0}\vee\bx (\Lamc )\Big)\right]\\
\quad\diy =\int_{\ocW^*_{\,\td}(\bx_0,\bx_0)}
{\obbP}^*_{\;\bx_0,\bx_0}({\rd}\oOm^*_0)\exp\;\left[-h\Big(\oOm^*_0+\orZ_0^*\big|
\Om^*_{\Lam\setminus\Lam_0}\vee\bx (\Lamc )\Big)\right]\\
\qquad\qquad\diy\times\prod\limits_{1\leq i\leq n}\exp\,
\Big\{- |x(i)-y(i)|^2_{\rm{Eu}}\big/[2k(\oom^*(i))\beta]\;\Big\}.\end{array}\eqno (3.3)$$
Here $\orZ_0^*$ is a collection of linear paths: $\orZ_0^*=(\ozeta^*(1),\ldots ,\ozeta^*(n))$
where
$$\ozeta^*(i):\ttt\in\big[0,k(\oom^*(i)\beta\big]\mapsto\frac{\ttt}{k(\oom^*)\beta}(y(i)-x(i)),\;\;1\leq i\leq n.$$ 

The first aforementioned contribution to the gradient
emerges when we differentiate the term $\exp\,
\Big\{- |x(j)-y(j)|^2_{\rm{Eu}}\big/[2k(\oom^*(j))\beta]\;\Big\}$, the second comes from differentiating
the term $\exp\;\left[-h\Big(\oOm^*_0+\orZ_0^*\big|
\Om^*_{\Lam\setminus\Lam_0}\vee\bx (\Lamc )\Big)\right]$. Consequently,
we obtain a uniform bound for the absolute value of the gradient 
$$\begin{array}{c}
\diy \left|\nabla_yF^{\Lam_0}_{\Lam |\bx (\Lamc )}(\bx_0,\by_0)\right|\leq 
2{\sqrt{d}}\,L_0\;v_0!\left(1\vee\sum\limits_{k\geq 1}
\orho^k\Big/(2\pi\beta k)^{d/2}\right)^{v_0}\\
\qquad\qquad\diy\times\left(\sum\limits_{k\geq 1}
\orho^k\Big/(2\pi\beta k)^{1+d/2}\right) \\
\qquad\diy +{\ov V}^{(1)}\;v_0!\;\beta\big(\tR^d/\td^d\big)\left(1\vee\sum\limits_{k\geq 1}
\orho^k\Big/(2\pi\beta k)^{d/2}\right)^{v_0}\\
\qquad\qquad\qquad\quad\diy\times\left(\sum\limits_{k\geq 1}
\orho^k\Big/\big[(2\pi )^{d/2}(\beta k)^{-1+d/2}\,\big]\right)\end{array}\eqno (3.4)$$
where we again used (1.1.4) and (1.1.19).

Hence, the family of RDMKs $\{F^{\Lam_0}_{\Lam |\bx (\Lamc )}\}$ is compact in 
space $C^0(\cC_\td(\Lam_0)\times\cC_\td(\Lam_0))$. Let $F^{\Lam_0}$ be a limit-point
as $\Lam\nearrow\bbR^d$. Then we have the Hilbert--Schmidt convergence 
$$\diy\lim_{k\to\infty}
\int\limits_{\cC(\Lam_0)\times \cC(\Lam_0)}{\rd}^{\Lam_0}\bx_0\,{\rd}^{\Lam_0}\by_0
\left[F^{\Lam_0}_{\Lam |\bx (\Lamc )}(\bx_0,\by_0)-F^{\Lam_0}(\bx_0,\by_0)
\right]^2=0.$$
Consequently, the RDM $R_{\Lam |\bx (\Lamc )}^{\Lam_0}$ in $\cH (\Lam_0)$
converges to the infinite-volume RDM $R^{\Lam_0}$ determined by the kernel
$F^{\Lam_0}$, in the 
Hilbert-Schmidt norm: 
$$\left\|R_{\Lam |\bx (\Lamc )}^{\Lam_0}-R^{\Lam_0}\right\|_{\rm{HS}}\to 0.\eqno (3.5)$$ 
As was mentioned, applying Lemma 1 from \cite{Su1}
(see also Lemma 1.5 from \cite{KS1}),
we obtain the  trace-norm convergence:
$$\left\|R_{\Lam |\bx (\Lamc )}^{\Lam_0}-R^{\Lam_0}\right\|_{\rtr}\to 0.\eqno (3.6)$$ 
Invoking a standard diagonal process implies that the 
sequence of states $\vphi_{\Lam |\bx (\Lamc )}$ is w$^*$-compact. 

Alongside with the above argument, one can establish that the PMs 
$\mu_{\Lam |\bx (\Lamc )}$ form a compact family as $\Lam\nearrow\bbR^d$.
More precisely, we would like to show that $\forall$ given cube
$\Lam_0$, the family of PMs $\mu_{\Lam |\bx (\Lamc )}^{\Lam_0}$ on 
$(\cW_\td^*(\Lam_0),\fW (\Lam_0))$ is compact. To this end, it suffices to check that 
the family $\{\mu_{\Lam |\bx (\Lamc )}^{\Lam_0}\}$
is tight as the Prokhorov theorem will then guarantee compactness.

Following an argument from \cite{KS1}, tightness is a consequence of two facts. 
\medskip

(a) The reference measure ${\rd}^{\Lam_0}\Om^*_0$ on $\cW^*_\td(\Lam_0)$ 
(see  Definition 2.1.2 (vii)) is supported by LCs
with the standard continuity modulus $\sqrt{2\epsilon\ln\,(1/\epsilon)}$.

(b) The PDF $f_{\Lam |\bx(\Lamc )} (\Om^*)
=\diy\frac{\mu_{\Lam |\bx (\Lamc )}(\rd\Om^*)}{{\rd}^\Lam\Om^*}$ (cf. (2.2.6))
is bounded from above by a constant similar to the RHS of (3.1).

As a result, the family of limit-point PMs $\{\mu^{\Lam_0}:\;\Lam_0\subset\bbR^d\}$
has the compatibility property and therefore satisfies the assumptions
of the Kolmogorov theorem. This implies that there exists a unique
PM $\mu$ on $(\cW^*_\td(\bbR^d),\fW (\bbR^d))$ such that the restriction of $\mu$ on
the sigma-algebra $\fW (\Lam_0)$ coincides with
$\mu^{\Lam_0}$. 

The fact that $\mu$ is an FK-DLR PM follows from the above construction.
Hence, each limit-point state $\vphi$
falls in class $\fF_+ (z,\beta )$. 
This completes the proof of Theorems 1.1 and 2.1.

\subsection*{Acknowledgments}
This work has been conducted under Grant 2011/20133-0 provided by 
the FAPESP, Grant 2011.5.764.35.0 provided by The Reitoria of the 
Universidade de S\~{a}o Paulo and Grant 2012/04372-7. The authors
express their gratitude to NUMEC and IME, Universidade de S\~{a}o Paulo,
Brazil, for the warm hospitality.

\end{document}